\documentclass[showpacs,prl,onecolumn,aps,superscriptaddress,preprintnumbers]{revtex4}
\usepackage{amsmath,amssymb}
\usepackage{epsfig}
\usepackage{graphicx}
\usepackage{mathrsfs}
\usepackage{amsmath}
\usepackage{amsfonts}
\def\slashchar#1{\setbox0=\hbox{$#1$}     		
   \dimen0=\wd0                                 	
   \setbox1=\hbox{/} \dimen1=\wd1               	
   \ifdim\dimen0>\dimen1                        	
      \rlap{\hbox to \dimen0{\hfil/\hfil}}      	
      #1                                        	
   \else                                        	
      \rlap{\hbox to \dimen1{\hfil$#1$\hfil}}   	
      /                                         	
   \fi}

\renewcommand{\vec}{\boldsymbol}
\newcommand{\beq}{\begin{equation}}
\newcommand{\eeq}{\end{equation}}
\newcommand{\bea}{\begin{eqnarray}}
\newcommand{\eea}{\end{eqnarray}}
\newcommand{\ba}{\begin{array}}
\newcommand{\ea}{\end{array}}

\def\eq#1{{Eq.~(\ref{#1})}}
\def\fig#1{{Fig.~\ref{#1}}}
\newcommand{\bas}{\bar{\alpha}_S}
\newcommand{\as}{\alpha_S}
\newcommand{\nn}{\nonumber}

\newcommand{\h}{\frac{1}{2}}

\newcommand{\ga}{\gamma}

\newcommand{\Lb}{\left(}
\newcommand{\Rb}{\right)}
\def\pom{{I\!\!P}}

\begin{document}

\title{ CGC/saturation approach for high energy soft interactions:
 $\mathbf{v_2}$ in proton-proton collisions}
\author{E. ~Gotsman}
\email{gotsman@post.tau.ac.il}
\affiliation{Department of Particle Physics, School of Physics and Astronomy,
Raymond and Beverly Sackler
 Faculty of Exact Science, Tel Aviv University, Tel Aviv, 69978, Israel}
 \author{ E.~ Levin}
\email{leving@post.tau.ac.il, eugeny.levin@usm.cl}
\affiliation{Department of Particle Physics, School of Physics and Astronomy,
Raymond and Beverly Sackler
 Faculty of Exact Science, Tel Aviv University, Tel Aviv, 69978, Israel}
 \affiliation{Departemento de F\'isica, Universidad T\'ecnica Federico Santa Mar\'ia, and Centro Cient\'ifico-\\
Tecnol\'ogico de Valpara\'iso, Avda. Espana 1680, Casilla 110-V, Valpara\'iso, Chile} 
 \author{  U.~ Maor}
\email{maor@post.tau.ac.il}
\affiliation{Department of Particle Physics, School of Physics and Astronomy,
Raymond and Beverly Sackler
 Faculty of Exact Science, Tel Aviv University, Tel Aviv, 69978, Israel}
\author{S. Tapia}
\email{sebastian.tapia.aray@cern.ch}
\affiliation{Departemento de F\'isica, Universidad T\'ecnica Federico Santa Mar\'ia, and Centro Cient\'ifico-\\
Tecnol\'ogico de Valpara\'iso, Avda. Espana 1680, Casilla 110-V, Valpara\'iso, Chile}

\date{\today}


\begin{abstract}
 In this paper we continue our program to construct a model for
 high energy soft interactions, based on the CGC/saturation 
approach. We demonstrate that  in  our model which describes  diffractive
 physics as well as  multi-particle production at high energy, 
 the density variation mechanism leads to the value of $v_2$  
 which
 is about  $60\% \div 70\%$ of the measured  $v_2$.  Bearing in mind that
  in CGC/saturation approach there are  two other  mechanisms 
present:
 Bose enhancement in the wave function and 
 local anisotropy,  we  believe that  the azimuthal long range
 rapidity correlations in proton-proton collisions  stem from the
 CGC/saturation physics, and not from  quark-gluon plasma production. 
 \end{abstract}
 
 \preprint{TAUP-3007/16}

\maketitle

\section{Introduction}
The large body of  experimental data  on soft interactions at
 high 
energy\cite{ALICE,ATLAS,CMS,TOTEM,ALICEI,CMSI,ATLASI,CMSMULT,ATLASCOR},
presently,
 cannot be described  in terms of theoretical 
  high energy QCD (see \cite{KOLEB} for a review).

 In this paper we continue our effort\cite{GLMNI,GLM2CH,GLMINCL,GLMCOR,GLMSP}
 to 
comprehend such interactions, by
  constructing  a model that incorporates  the
 advantages of two theoretical approaches to high energy QCD: the
  CGC/saturation approach \cite{GLR,MUQI,MV,B,MUCD,K,JIMWLK} and the BFKL
 Pomeron
 calculus\cite{BFKL,LIREV,GLR,MUPA,BART,BRN,KOLE,LELU,LMP,AKLL,LEPP} .
 Both provide an effective theory of QCD at high energies. However,
  the interpretation of processes at high energy appear quite different in
 each,
  since  they have different  structural elements.
   
 The CGC/saturation approach  describes the high energy interactions
 in terms of colorless dipoles, their density, 
 distribution   over impact parameters, evolution in energy and so on.
 Such a description  appears quite natural in perturbative QCD, and can be
    easily applied  to 
   multi-particle production at high energy. In this approach a new 
saturation
  scale
      $Q_s$ occurs which  is much 
larger than the
 soft scale,  and the description of the production of quarks and gluons  
 can be  attained theoretically, in an economical way. The  
transition from 
quarks
 and gluons to hadrons,  has to be handled phenomenologically  using data 
from  
hard
 processes. This approach leads to a good description of the 
experimental
 data on inclusive production, and the observation of some regularities in 
the data,
 such as  geometric scaling behaviour\cite{KLN,KLNLHC,LERE,LERECOR,MCLPR,PRA}.

 BFKL Pomeron calculus   which  deals with   BFKL Pomerons and
 their  
interactions, is similar to the old  Reggeon theory \cite{COL}, and is 
suitable for   describing 
 diffractive physics and correlations in multi-particle production,
 as we can use
 the Mueller diagram technique\cite{MUDI}.   
  The relation
 between these two approaches has not yet been established,  but
 they are equivalent\cite{AKLL} for  rapidities ($\ln \Lb s/s_0\Rb$), 
such that 
\beq \label{I1}
Y \,\leq\,\frac{2}{\Delta_{\mbox{\tiny BFKL}}}\,\ln\Lb
 \frac{1}{\Delta^2_{\mbox
{\tiny BFKL}}}\Rb
\eeq
where $\Delta_{\mbox{\tiny BFKL}}$ denotes the intercept of the BFKL 
Pomeron.
 As we have discussed \cite{GLMNI}, the parameters of our model are
  such that for $Y\, \leq\,36$,
 we can  trust our approach, based on the BFKL 
Pomeron calculus.

 This paper is the next step in our program to construct a model for high
 energy soft scattering, based on an analytical calculation, without 
 using  Monte Carlo simulations.
 
 One of the most intriguing experimental observations made at the LHC
 and RHIC, is the same pattern of  azimuthal angle correlations in
the  three types of  interactions: hadron-hadron, hadron-nucleus and
 nucleus-nucleus collisions. In all three reactions,  correlations
in the events with large density of produced particles,
 are observed between two charged hadrons, which are separated by the
 large values of rapidity.
  \cite{CMSPP,STARAA,PHOBOSAA,STARAA1,CMSPA,CMSAA,ALICE}. 

We believe that the results of these experiments provide  strong evidence,
 that the underlying physics is the same for all three reactions,
 and  is due the high density partonic state  that has
 been produced at high energies, in all three reactions. Due to
 causality arguments\cite{CAUSALITY} two hadrons with large difference
 in rapidity between them, could only correlate  at  the early stage 
 of the collision and, therefore,  we expect that the correlations
 between two particles with large rapidity difference (at least the
 correlations in rapidity) are due to the partonic state with large
 parton density.  
The parton (gluon) density is governed in QCD by the CGC/saturation
 non-linear equations and therefore,
 the CGC approach is the appropriate  tool to study these correlations.

Unlike, the rapidity correlations at large values of the rapidity
 difference, which stem from the initial state interactions, the
 azimuthal angle correlations can  originate from the collective
 flow in the final state \cite{FINSTATE}.  Nevertheless, in this paper
 we would like to analyze the same mechanism for both correlations: i.e. 
the
 initial state interaction in the CGC phase of QCD. However, even in the
 framework of the saturation/CGC approach, we presently are not able to
propose
 a unique mechanism for the azimuthal angle correlations.   At the moment
   three sources for the azimuthal angle correlation have been suggested:
\footnote{We use the classification and terminology suggested in Ref. 
\cite{KOVT}.
}
Bose enhancement in the wave function\cite{DDGJLR}, 
 local anisotropy\cite{KOLUCOR,KOLUREV} and density variation \cite{LERECOR}.
 We cite only a restricted number of papers for each approach. The reader 
can
 find more references, and more ideas on the origin of the correlations in 
the
 review papers  Refs.\cite{KOLUREV,REV1,REV2,REV3,REV4,REV5,REV6}.

The goal of this paper  is to study in more detail the density 
variation
 mechanism proposed in Ref.\cite{LERECOR}. In this approach,  both 
rapidity
 and  azimuthal angle correlations originate from two gluons production  
from
 two parton showers.

 This process can be  calculated  using Mueller diagrams \cite{MUDI} 
 (see \fig{nimcor2sh}-a).  The difference between rapidity and
 azimuthal angle correlations is only in the form of the Mueller
 vertices in this figure.
For rapidity correlations, such vertices can be considered as being
 independent of $Q_T$, while for the
azimuthal angle correlation, this vertex is proportional to
 $\Lb\vec{Q}_T \cdot p_{1,\bot}\Rb^2$ or $
\Lb\vec{Q}_T \cdot p_{2,\bot}\Rb^2$. The integration over
 the direction of $\vec{Q}_T$  leads to the term
 $\Lb \vec{p}_{1,\bot}\cdot \vec{p}_{2 \bot}\Rb^2$,
 which is proportional to $\cos 2 \varphi$, resulting
 in  azimuthal angle correlations. The strength of
 the term  $\cos 2 \varphi$ is proportional to $\langle
 Q_T^2\rangle^2$, where averaging is  over the wave
 function of the one parton shower, which is described by the
 BFKL Pomeron. In other words since $\vec{Q}_T = i \nabla_b$,
 where $b$ is the impact factor for the scattering process,
 the magnitude of the azimuthal angle correlation depends
 on the gradient of the parton density.

  
  \section{The brief review of the model}
  In this section we briefly review  our model, which provides a
successful  description of the diffractive\cite{GLMNI,GLM2CH}, inclusive
 cross sections\cite{GLMINCL} and rapidity correlations\cite{GLMCOR}.
  For the description of the angular correlations, it is
 important to take into account  the dependance of  the 
scattering
 amplitude on the sizes of dipoles,  hence in all formulae below, we
 include this dependance. In our description of the cross sections
 we previously took $r = 1/m$.
 
\begin{figure}[ht]
    \centering
  \leavevmode
      \includegraphics[width=14cm]{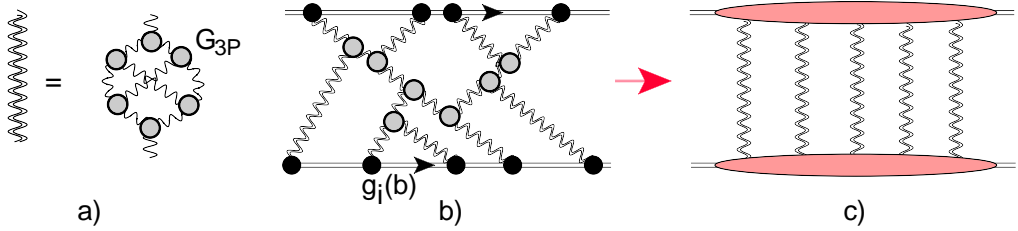}  
      \caption{\fig{amp}-a shows the set of the diagrams in the
 BFKL Pomeron calculus that produce the resulting (dressed) Green
 function of the Pomeron in the framework of high energy QCD.
 In \fig{amp}-b the net diagrams    which   include
 the interaction of the BFKL Pomerons with colliding hadrons are shown.
 The sum of the diagrams reduces to \fig{amp}-c after integration
 over positions of $G_{3 \pom}$ in rapidity. }
\label{amp}
   \end{figure}

 The main ingredient of our model is
 the BFKL Pomeron  Green function, which we determined using the 
CGC/saturation
 approach\cite{GLMNI,LEPP}. We calculated this function  from the 
solution of
 the non-linear Balitsky-Kovchegov equation\cite{B,K},  using the MPSI
 approximation\cite{MPSI} to sum enhanced diagrams shown in \fig{amp}-a.
 It has the following form:
 \bea \label{G}
G^{\mbox{\tiny dressed}}\Lb T\Rb\,\,&=&\,\,a^2 (1 - \exp\Lb -T\Rb )  +
 2 a (1 - a)\frac{T}{1 + T} + (1 - a)^2 G\Lb T\Rb \nn\\
~~~&\mbox{with}&~~G\Lb T\Rb = 1 - \frac{1}{T} \exp\Lb \frac{1}{T}\Rb
 \Gamma_0\Lb \frac{1}{T}\Rb
\eea

\beq \label{T}
T\Lb r_\bot, s, b\Rb\,\,=\,\,\phi_0  \Lb r^2_\bot Q^2_s\Lb Y, b\Rb\Rb^{\bar \gamma}  
\eeq
where the saturation momentum $Q_s$ is  given by
\beq \label{QS}
Q^2_s\Lb b, Y\Rb\,\,=\,\,Q^2_{0s}\Lb b, Y_0\Rb\,e^{\lambda \,(Y - Y_0)}
\eeq
with
\beq \label{QS0}
Q^2_{0s}\Lb b, Y_0\Rb\,\,=\,\, \Lb m^2\Rb^{1 - 1/\bar \gamma}\,\Lb S\Lb b,
 m\Rb\Rb^{1/\bar{\gamma}} 
~~~~~~~S\Lb b , m \Rb \,\,=\,\,\frac{m^2}{2 \pi} e^{
 - m b}~~~\mbox{and}~~\bar \gamma\,=\,0.63
\eeq 
 $T\Lb r_\bot, s, b\Rb$ denotes the dipole - proton amplitude
 in the vicinity of the saturation scale \cite{MUTR}.

 In these formulae we take $a=0.65$,  this value was chosen,  to 
reproduce
 the analytical form for the solution of the BK equation. Parameters
 $\lambda$ and $\phi_0$, can be estimated in the leading order of  QCD,
 but due to large next-to-leading order corrections, we treat 
them as parameters of  the fit. $m$ is a non-perturbative parameter,
 which characterizes the large impact parameter behavior of the
 saturation momentum, as well as the typical size of dipoles that
 take part in the interactions. The value of $m =5.25\,GeV$ in our
 model, justifies our main assumption, that  BFKL Pomeron calculus
 based on a perturbative QCD approach, is able to describe  soft 
physics,
 since $m \,\gg\,\mu_{soft}$, where $\mu_{soft}$ denotes the natural scale 
for
 soft processes ($ \mu_{soft} \,\sim\,\Lambda_{QCD}$ and/or  pion mass).
 
 Unfortunately, since the confinement problem is far from being 
solved, we assume
 a phenomenological approach for the structure of the colliding hadrons.
 We use a two channel model, which allows us to calculate the
 diffractive production in the region of small masses.
   In this model, we replace the rich structure of the 
 diffractively produced states, by a single  state with the wave 
function 
$\psi_D$, a la Good-Walker\cite{GW}.
  The observed physical 
hadronic and diffractive states are written in the form 
\beq \label{MF1}
\psi_h\,=\,\alpha\,\Psi_1+\beta\,\Psi_2\,;\,\,\,\,\,\,\,\,\,\,
\psi_D\,=\,-\beta\,\Psi_1+\alpha \,\Psi_2;~~~~~~~~~
\mbox{where}~~~~~~~ \alpha^2+\beta^2\,=\,1;
\eeq 

Functions $\psi_1$ and $\psi_2$  form a  
complete set of orthogonal
functions $\{ \psi_i \}$ which diagonalize the
interaction matrix ${\bf T}$
\beq \label{GT1}
A^{i'k'}_{i,k}=<\psi_i\,\psi_k|\mathbf{T}|\psi_{i'}\,\psi_{k'}>=
A_{i,k}\,\delta_{i,i'}\,\delta_{k,k'}.
\eeq
The unitarity constraints take  the form
\beq \label{UNIT}
2\,\mbox{Im}\,A_{i,k}\left(s,b\right)=|A_{i,k}\left(s,b\right)|^2
+G^{in}_{i,k}(s,b),
\eeq
where $G^{in}_{i,k}$ denotes the contribution of all non 
diffractive inelastic processes,
i.e. it is the summed probability for these final states to be
produced in the scattering of a state $i$ off a state $k$. In \eq{UNIT} 
$\sqrt{s}=W$ denotes the energy of the colliding hadrons, and $b$ 
the 
impact  parameter.
A simple solution to \eq{UNIT} at high energies, has the eikonal form 
with an arbitrary opacity $\Omega_{ik}$, where the real 
part of the amplitude is much smaller than the imaginary part.
\beq \label{A}
A_{i,k}(s,b)=i \Lb 1 -\exp\Lb - \Omega_{i,k}(s,b)\Rb\Rb,
\eeq
\beq \label{GIN}
G^{in}_{i,k}(s,b)=1-\exp\Lb - 2\,\Omega_{i,k}(s,b)\Rb.
\eeq
\eq{GIN} implies that $P^S_{i,k}=\exp \Lb - 2\,\Omega_{i,k}(s,b) \Rb$, is 
the probability that the initial projectiles
$(i,k)$  reach the final state interaction unchanged, regardless of 
the initial state re-scatterings.
\par

Note, that there is no factor $1/2$, its absence stems from our definition
 of the dressed Pomeron.


\begin{table}[h]
\begin{tabular}{|l|l|l|l|l|l|l|l|l|l|}
\hline
model &$\lambda $ & $\phi_0$ ($GeV^{-2}$) &$g_1$ ($GeV^{-1}$)&$g_2$ ($GeV^{-1}$)& $m(GeV)$ &$m_1(GeV)$& $m_2(GeV)$ & $\beta$& $a_{\pom \pom}$\\
\hline
 2 channel & 0.38& 0.0019 & 110.2&  11.2 & 5.25&0.92& 1.9 & 0.58 &0.21 \\
\hline
\end{tabular}
\caption{Fitted parameters of the model. The values are taken 
from Ref.\cite{GLM2CH}.}
\label{t1}
\end{table}
In the eikonal approximation we replace $ \Omega_{i,k}(r_\bot, s,b)$ by 
\beq \label{EAPR}
 \Omega_{i,k}(r_\bot, Y - Y_0,b)\,\,=\,\int d^2 b'\,d^2 b''\,
 g_i\Lb \vec{b}'\Rb \,G^{\mbox{\tiny dressed}}\Lb T\Lb r_\bot,
 Y - Y_0, \vec{b}''\Rb\Rb\,g_k\Lb \vec{b} - \vec{b}'\ - \vec{b}''\Rb 
 \eeq
 We propose a more general approach, which takes into account new
 small parameters, that come from the fit to the experimental data
 (see Table 1 and \fig{amp}):
 \beq \label{NEWSP}
 G_{3\pom}\Big{/} g_i(b = 0 )\,\ll\,\,1;~~~~~~~~ m\,\gg\, m_1 
~\mbox{and}~m_2
 \eeq
 
 The second equation in \eq{NEWSP} leads to the fact that $b''$ in 
\eq{EAPR} is much
 smaller than $b$ and $ b'$,
  therefore, \eq{EAPR} can be re-written in
 a simpler form
 \bea \label{EAPR1}
 \Omega_{i,k}(r_\bot, Y - Y_0, b)\,\,&=&\,\Bigg(\int d^2 b''\,
G^{\mbox{\tiny dressed}}\Lb
 T\Lb r_\bot, Y - Y_0, \vec{b}''\Rb\Rb\Bigg)\,\int d^2 b' g_i\Lb
 \vec{b}'\Rb \,g_k\Lb
 \vec{b} - \vec{b}'\Rb \,\nn\\
 &=&\,\tilde{G}^{\mbox{\tiny dressed}}\Lb r_\bot, Y - Y_0\Rb\,\,
\int d^2 b' g_i\Lb
 \vec{b}'\Rb \,g_k\Lb \vec{b} - \vec{b}'\Rb 
\eea

  Selecting the diagrams using the first equation in \eq{NEWSP}, indicates
 that the main contribution stems from the net diagrams shown in \fig{amp}-b.
 The sum of these diagrams\cite{GLM2CH} leads to the following expression 
for $
 \Omega_{i,k}(s,b)$
 \bea
\Omega\Lb r_\bot,  Y-Y_0; b\Rb~~&=& ~~ \int d^2 b'\,
\,\,\,\frac{ g_i\Lb\vec {b}'\Rb\,g_k\Lb\vec{b} -
 \vec{b}'\Rb\,\tilde{G}^{\mbox{\tiny dressed}}\Lb r_\bot, Y - Y_0\Rb
}
{1\,+\,G_{3\pom}\,\tilde{G}^{\mbox{\tiny dressed}}\Lb r_\bot, Y - Y_0\Rb\left[
g_i\Lb\vec{b}'\Rb + g_k\Lb\vec{b} - \vec{b}'\Rb\right]} ;\label{OM}\\
g_i\Lb b \Rb~~&=&~~g_i \,S_p\Lb b; m_i \Rb ;\label{g}
\eea
where
\beq \label{SB}
S_p\Lb b,m_i\Rb\,=\,\frac{1}{4 \pi} m^3_i \,b \,K_1\Lb m_i b \Rb
\eeq
$$
\tilde{G}^{\mbox{\tiny dressed}}\Lb r_\bot, Y -Y_0\Rb\,\,=\,\,\int d^2 b
 \,\,G^{\mbox{\tiny dressed}}\Lb T\Lb r_\bot, Y - Y_0, b\Rb\Rb
$$
where $ T\Lb r_\bot, Y - Y_0, b\Rb$ is given by \eq{T}.

Note  that  $\bar{G}^{\mbox{\tiny dressed}}\Lb \bar T\Rb$ does not depend
 on $b$.  In all previous formulae, the value of the triple BFKL Pomeron vertex
 is known: $G_{3 \pom} = 1.29\,GeV^{-1}$.

To simplify further discussion, we introduce the notation 

 \beq \label{NBK}
N^{BK}\Lb G^i_\pom\Lb r_\bot, Y,b \Rb\Rb \,\,=\,\,a\,\Lb 1
 - \exp\Lb -  G^i_\pom\Lb r_\bot, Y, b\Rb\Rb\Rb\,\,+\,\,\Lb 1 - a\Rb
\frac{ G^i_\pom\Lb  r_\bot, Y, b\Rb}{1\,+\, G^i_\pom\Lb r_\bot, Y, b\Rb},
\eeq 
 with $a = 0.65$ .
 \eq{NBK} is an analytical approximation to the numerical solution for  
the 
BK equation\cite{LEPP}. $G_\pom\Lb  r_\bot, Y; b\Rb \,=\,\,
 g_i\Lb b \Rb \,\tilde{G}^{\mbox{\tiny dressed}}\Lb r_\bot, Y - Y_0\Rb $.
 We recall that the BK equation sums the `fan'  diagrams.

For the  elastic amplitude we have

\beq \label{EL}
a_{el}(b)\,=\,\Lb \alpha^4 A_{1,1}\,
+\,2 \alpha^2\,\beta^2\,A_{1,2}\,+\,\beta^4 A_{2,2}\Rb. 
\eeq

\section{Correlations between two parton showers}

\begin{figure}[ht]
    \centering
  \leavevmode
      \includegraphics[width=15cm]{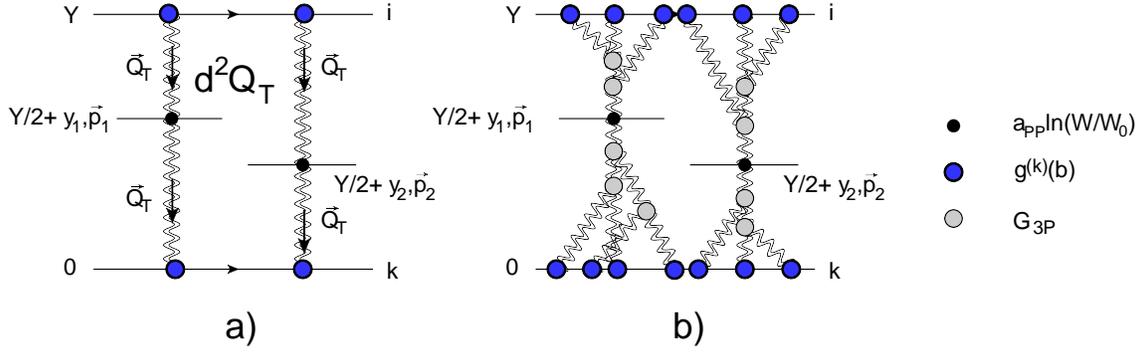}  
      \caption{The  Mueller diagram for the  rapidity correlation between
 two particles produced in two parton showers. \protect\fig{nimcor2sh}-a  shows
 the first Mueller diagram, while\protect\fig{nimcor2sh}-b indicates the 
structure
 of general diagrams.
  The double  wavy lines describe the dressed  BFKL Pomerons.
 The blobs stand for the vertices as shown in the legend.}
\label{nimcor2sh}
   \end{figure}

In our previous paper \cite{GLMCOR}, we discovered that in the framework
 of our model that has been described above, the main source of the long
 range rapidity correlation, is the correlation  between two parton 
showers.
  The appropriate   Mueller diagrams are shown
 in \fig{nimcor2sh}. 
 Examining this diagram, we see that the contribution to the double
 inclusive cross section, differs from the product of two single inclusive
 cross sections.  This difference generates the rapidity correlation
 function, which is defined as
 \beq \label{RCF}
  R\Lb y_1, y_2\Rb\,\,=\,\ \frac{\frac{1}{\sigma_{in}} \frac{d^2 \sigma}{d y_1\,d y_2}}{\frac{1}{\sigma_{in}} \frac{d \sigma}{d y_1} \, \frac{1}{\sigma_{in}} \frac{d \sigma}{d y_2}  }\,\,-\,\,1
  \eeq

     \begin{figure}[ht]
    \centering
  \leavevmode
      \includegraphics[width=5cm]{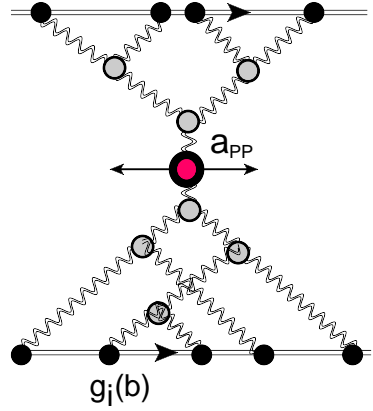}  
      \caption{ The Mueller diagram for the  single inclusive cross 
section.
  The  double wavy lines describe  the resulting Green function of
 the BFKL Pomerons ( $\tilde{G}^{\mbox{\tiny dressed}} $). The blobs
 stand for the vertices which are  the same as in \fig{nimcor2sh}.}
\label{incl}
   \end{figure}


  There are two reasons for the difference between the 
 double inclusive cross section due to production of 
 two parton showers, and the products of inclusive cross
 sections: the first, is that in the 
expression for the double
 inclusive cross section, we integrate  the product of the single
 inclusive inclusive cross sections, over $b$. The second, 
is that the 
summation 
 over $i$ and $k$  for  the product of single  inclusive cross
 sections, is for fixed $i$ and $k$.

Introducing the  following new function, enables us to write the 
 analytical expression for the double inclusive cross section: 
\bea \label{2SH1}
&&I^{(i,k}\Lb y, b \Rb\,\,=\,\,a_{\pom \pom}\,\ln\Lb W/W_0\Rb\\
&& \times\int d^2 b' \,\,N^{BK}\Bigg( g^{(i)}\,S\Lb m_i,b'\Rb
 \tilde{G}^{\mbox{\tiny dressed}}\Lb r_\bot = 1/m, \h Y + y
 \Rb\Bigg)\,\,\nn\\
 && N^{BK}\Bigg( g^{(k)}\,S\Lb m_k,
 \vec{b} - \vec{b}^{\,'}\Rb \tilde{G}^{\mbox{\tiny dressed}}\Lb
 r_\bot=1/m,\h Y- y \Rb\Bigg)\nn
\eea

 Using \eq{2SH1} we can write the double inclusive cross section in the form
 \bea \label{2SH2}
 &&\frac{d^2 \sigma^{\mbox{\tiny 2 parton showers}}}{ d
 y_1\,\,d y_2}\,\,=\,\,\int d^2p_{1T}\,d^2 p_{2T}\frac{d^2 \sigma^{\mbox{\tiny 2 parton showers}}}{ d
 y_1\,\,d y_2\,d^2 p_{1T}\,d^2 p_{2T}}\,\,=\,\,\int d^2 b \,\,\Bigg\{ \alpha^4
 \,I^{(1,1)}\Lb y_1, b \Rb\,I^{(1,1)}\Lb y_2, b \Rb\nn\\
 && \alpha^2\,\beta^2\,\Lb  I^{(1,2)}\Lb y_1, b \Rb
\,I^{(1,2)}\Lb y_2, b \Rb\,\,+\,\,I^{(2,1)}\Lb y_1,
 b \Rb\,I^{(2,1)}\Lb y_2, b \Rb \Rb\,\,+\,\,\beta^4\,I^{(2,2)}\Lb
 y_1, b \Rb\,I^{(2,2)}\Lb y_2, b \Rb \Bigg\}
 \eea

 Comparing \eq{2SH2} with   the square of the single inclusive
 cross section (see  below  \eq{INCF}) , we note the
 different powers of $\alpha$ and $\beta$, which reflect the different
 summation over $i$ and $k$, as well as different integration over $b$.
 
 Other sources can contribute to the correlation function
 $R\Lb y_1,y_2\Rb$ which is defined as
 \beq \label{R1}
 R\Lb y_1,y_2\Rb\,\,=\,\,\sigma_{NSD}\,\Bigg\{
  \frac{d^2 \sigma^{\mbox{\tiny 2 parton showers}}}
{ d y_1\,\,d y_2}\,\,+\,\,\frac{d^2 \sigma^{\mbox{\tiny 1
 parton shower}}_{\mbox{\tiny semi-enhanced}}}{ d y_1\,\,d y_2}\,\,
+\,\, \frac{d^2 \sigma^{\mbox{\tiny 1 parton shower}}_{\mbox{\tiny enhanced}}}
{ d y_1\,\,d y_2}\Bigg\}\Bigg{/}
 \Bigg\{ \frac{d \sigma}
 {d y_1}\,\frac{d \sigma}
 {d y_2}\Bigg\}\,\,\,-\,\,\,1
 \eeq
 
  In Ref.\cite{GLMCOR} we showed that both 
semi-enhanced and enhanced diagrams which are related to the
 correlations in one patron shower, give negligible contributions,
 and can be neglected.
 
 We have discussed in Ref.\cite{GLMCOR}
 the rapidity correlations that are generated by \eq{2SH2}. 
 In the present paper we  wish to consider
 the correlations in the azimuthal angle between two momenta of 
 produced gluons: $\vec{p}_{1,\bot}$ and $\vec{p}_{2, \bot}$.
 
  The single inclusive cross section can be calculate using the following
 formula\cite{GLMINCL}
 \bea \label{INCF}
   \frac{d \sigma}{d y}\,\,&=&\,\,\int d^2 p_T\,\frac{d \sigma}{d y
 \,d^2 p_T}\,\,=\,\,a_{\pom \pom}\,\ln\Lb W/W_0\Rb\Bigg\{ \alpha^4 
 \,In^{(1)}\Lb \h Y + y\Rb \, In^{(1)}\Lb \h Y - y\Rb  \,\nn\\
   &+&\,\alpha^2\beta^2 \Big(In^{(1)}\Lb \h Y + y\Rb \,
 In^{(2)}\Lb \h Y - y\Rb \,+\,  In^{(2)}\Lb \h Y + y\Rb \,
 In^{(1)}\Lb \h Y - y\Rb\Big)\,\nn\\
   &+&\,\beta^4 \,In^{(2)}\Lb \h Y + y\Rb \, In^{(2)}\Lb \h Y - y\Rb  \Bigg\}
\eea
where $Y$ denotes the total rapidity of the colliding particles,  and $y$
 is the rapidity of produced hadron. $In^{(i)}(y)$ is given by
\beq \label{IN}
In^{(i)}\Lb y\Rb\,=\,\int d^2 b \,\,N^{BK}\Lb g^{(i)}\,S\Lb m_i,
 b\Rb \,\tilde{G}_\pom\Lb r_\bot = 1/m, Y - Y_0\Rb\Rb
\eeq
and 
$a_{\pom\pom}$ is a fitted parameter, that was determined
 in Ref.\cite{GLMINCL} (see Table 1).

 \section{Calculation of the first diagram}
 In this section we calculate the first Mueller diagram shown
 in \fig{nimcor2sh}-a.  We reproduce in an alternative
 way, the main results of Ref. \cite{LERECOR}. We start from the
 calculation of the inclusive production, from one BFKL Pomeron,
  which enters  this diagram at fixed momentum transfer  $\vec{Q}_T$.
  Note, that the inclusive cross section, shown in \fig{incl},
 is determined by the same BFKL Pomeron, but at $Q_T = 0$.
 
 \subsection{Inclusive production from one BFKL Pomeron}
  \subsubsection{The BFKL Pomeron: generalities}
The general solution to the BFKL equation for the scattering amplitude of
 two dipoles with the sizes $r_1$ and $r_2$, has been derived in
 Ref.\cite{LIREV}, and has the form
\bea \label{GENSOL}
&&N_\pom\Lb r_1,r_2; Y, b \Rb\,\,= \\
&&\,\,\sum_{n=0}^{\infty}
\int \frac{d \gamma}{2\,\pi\,i}\,\phi^{(n)}_{in}(\gamma; r_2)
\,\,d^2\, R_1 \,\,d^2\,R_2\,\delta(\vec{R}_1 - \vec{R}_2 - 
\vec{b})\,
e^{\omega(\gamma, n )\,Y}
\,E^{\gamma,n}\Lb r_1, R_1\Rb\,E^{1 - \gamma,n}\Lb r_2, R_2 \Rb\nn
\eea
with
\beq \label{OMEGA}
\omega(\gamma, n)\,\,=\,\,\bas \chi(\gamma, n)\,\,  =\,\,\bas \Lb 2 
\psi\Lb 1\Rb \,-\,\psi\Lb \gamma + |n|/2\Rb\,\,-\,\,\psi\Lb 1 
 - \gamma + |n|/2\Rb\Rb;
\eeq
where $~\psi\Lb \gamma\Rb \,\,=\,\,d \ln \Gamma\Lb \gamma\Rb/d
 \gamma $ and $\Gamma\Lb \gamma\Rb$ is  Euler gamma function. 
Functions $ E^{n, \gamma} \Lb \rho_{1a},\rho_{2a}\Rb$ are given
 by the following equations.
\begin{align}\label{EFUN}
  E^{n, \gamma} \Lb \rho_{1a},\rho_{2a}\Rb \,=\, \Lb
  \frac{\rho_{12}}{\rho_{1a} \, \rho_{2a}}\Rb^{1 - \gamma + n/2}
  \, \Lb \frac{\rho^*_{12}}{\rho^*_{1a} \, \rho^*_{2a}}
  \Rb^{1 - \gamma - n/2},
\end{align}
In \eq{EFUN} we use  complex numbers to characterize the point
 on the plane
\begin{align}\label{COMNUM}
  \rho_i = x_{i,1} + i \, x_{i,2};\,\,\,\,\,\,\, \rho^*_i = x_{i,1} - 
 i \, x_{i,2}
\end{align}
where the indices $1$ and $2$ denote  two transverse axes. Notice that
\beq \label{NOT}
\rho_{12}\,\rho^*_{12}\,\,=\,\,r^2_i ;~~~~~~\rho_{1 a}\,\rho^*_{1 a}\,=
\,\Lb\vec{R}_i\,-\,\frac{1}{2}\vec{r}_{i}\Rb^2~~~~~~\rho_{2 a}\,\rho^*_{2a}\,
=\,\Lb\vec{R}_i\,+\,\frac{1}{2}\vec{r}_{i}\Rb^2
\eeq
At large values of $Y$, the main contribution stems from the first term 
with $n =0$.  For this term \eq{EFUN} can be re-written in the form
\beq \label{E}
E^{\gamma,0}\Lb r_i,R_i\Rb \,\,=\,\,\left( \,\frac{r^2_{i}}{(\vec{R}_i
\,+\,\frac{1}{2}\vec{r}_{i})^2\,\,
(\vec{R}_i\,-\,\frac{1}{2}\vec{r}_{i})^2}\,\right)^{1 - \gamma}\,\,.
\eeq

The integrals over $R_1$ and $R_2$ were taken in Refs.\cite{LIREV,NAPE}
 and at $n=0$ we have
\bea \label{H}
&&H^\ga\Lb w, w^*\Rb\,\,\equiv\,\,\int d^2\,R_1\,E^{\ga,0}\Lb r_{1},R_1\Rb\,
 E^{1 - \ga, 0}\Lb r_{2}, \vec{R}_1
\,-\,\vec{b}\Rb\,= \\
&&
\,\frac{ (\gamma - \h)^2}{( 
\gamma (1 - \gamma)
)^2} \Big\{b_\ga\,w^\gamma\,{w^*}^\gamma\,F\Lb\gamma, \gamma, 2\gamma, w\Rb\,
F\Lb\gamma, \gamma, 2\gamma, w^*\Rb
\,+ \nn\\
&&  b_{1 - \ga} w^{1 -
\gamma}{w^*}^{1-\gamma}
F\Lb 1 - \gamma, 1 -\gamma, 2 - 2\gamma, w\Rb\,F\Lb 1 - \gamma,1 -\gamma,2 
-2\gamma, w^*\Rb \Big\}\nn
\eea
where $F$ is hypergeometric function \cite{RY}. In  \eq{H}
$w\,w^*$ is equal to
\beq \label{W}
w\,w^*\,\,=\,\,\frac{r^2_{1}\,r^2_{2}}{\Lb\vec{b} - \h\Lb\,\vec{r}_{1}\,
- \,\vec{r}_{2}\Rb\Rb^2
\,\Lb\vec{b} \,+\, \h \Lb\,\vec{r}_{1} \,- \,\vec{r}_{2}\Rb\Rb^2}
\eeq
and  $b_\ga$ is equal to
\beq \label{BGA}
b_{\ga} \, = \, \pi^3 \, 2^{4(1/2 - \ga)} \, \frac{\Gamma \Lb\ga \Rb}{\Gamma \Lb 1/2 - \ga \Rb}
  \, \frac{\Gamma \Lb 1 - \ga  \Rb}{\Gamma \Lb 1/2 + \ga \Rb}.
\eeq

Finally, the solution at large $Y$ has the form
\beq \label{FINSOL}
N_\pom\Lb r_1,r_2; Y, b \Rb\,\,= \,\,
\int \frac{d \gamma}{2\,\pi\,i}
\,
e^{\omega(\gamma, 0 )\,Y}\,H^\ga\Lb w, w^*\Rb
\eeq

In the vicinity of the saturation scale $N_\pom$ takes the form
 (see Refs. \cite{MUTR,IIMU})
\bea 
&&N_\pom\Lb r_1,r_2; Y, b \Rb\,\,= \,\,\frac{\Lb \ga_{cr} - \h\Rb^2}{\ga_{cr} ( 1 - \ga_{cr})}\,b_{\ga_{cr}}\Big(w w^* e^{\kappa Y})^{1 - \gamma_{cr}}\nn\\
&&\,\,=\,\,\frac{\Lb \ga_{cr} - \h\Rb^2}{\ga_{cr} ( 1 - \ga_{cr})}\,b_{\ga_{cr}}\Lb \Lb \frac{r^2_{1}\,r^2_{2}}{\Lb\vec{b} - \h\Lb\,\vec{r}_{1}\,
- \,\vec{r}_{2}\Rb\Rb^2
\,\Lb\vec{b} \,+\, \h \Lb\,\vec{r}_{1} \,- \,\vec{r}_{2}\Rb\Rb^2}\Rb e^{\bas \frac{\chi\Lb \ga_{cr}\Rb}{1 - \ga_{cr}}\,Y}\Rb^{1 - \gamma_{cr}} \label{GSNB}\\
&&\xrightarrow{r_2 \gg r_1}~~~~ \phi_0\,\Big( r^2_1 Q^2_s\Lb r_2,b; Y\Rb\Big)^{1 - \gamma_{cr} }~~~~\mbox{with}~~~~~Q^2_s\Lb r_2,b; Y\Rb\,\,= \frac{\,r^2_{2}\,e^{\bas \frac{\chi\Lb \ga_{cr}\Rb}{1 - \ga_{cr}}\,Y}}{\Lb\vec{b} - \h\vec{r}_{2}\Rb^2
\,\Lb\vec{b} \,+\, \h \vec{r}_{2}\Rb^2}\label{GSNB1}
\eea
where (see Refs.\cite{GLR,MUTR,MUPE})
\beq \label{GACR}
\frac{\chi\Lb \ga_{cr}\Rb}{1 - \ga_{cr}}\,\,=\,\,- \frac{d \chi\Lb \ga_{cr}\Rb}{d \ga_{cr}}
\,\,\,\,\,\mbox{where}\,\,\,\,\,\,\chi\Lb \ga\Rb\,=\,2 \psi\Lb 1 \Rb \,-\,\psi\Lb \ga\Rb \,-\,\psi\Lb 1 - \ga\Rb \,\leftarrow \mbox{kernel of the BFKL equation}
\eeq
  Below we denote   by $\bar{\gamma} = 1 - \gamma_{cr}$, and  will use
 \eq{GSNB} and \eq{GSNB1} in the  momentum transfer   
representation, viz.
\beq \label{QRE}
N_\pom\Lb r_1,r_2; Y, Q_T \Rb\,\,\,=\,\,\int d^2 b \,\,e^{i
 \vec{Q}_T \cdot \vec{b}}\,\,\,N_\pom\Lb r_1,r_2; Y, b \Rb\,
\eeq

The integral of \eq{QRE} with $ N_\pom\Lb r_1,r_2; Y, b \Rb$ from 
\eq{GSNB}
  can be  evaluated using the complex number description for the 
point on the
 plane, (see \eq{COMNUM} and \eq{NOT}).   The integral has the
 form\cite{LIREV,NAPE}
\beq \label{NPQ10}
N_\pom\Lb r_1,r_2; Y, Q_T \Rb =\,\Lb r^2_1\,r^2_2 \Rb^{\bar{\gamma}}
 \,e^{\bas \, \chi\Lb \ga_{cr}\Rb\,Y}\int d \rho_b  \,\,e^{i \rho^*_Q 
\rho_b}\Lb  \frac{1}{\rho^2_b - \rho^2_{12}}\Rb^{\gamma}\int d \rho^*_b 
 \,\,e^{i \rho_Q \rho^*_b} \Lb \frac{1}{\rho^{*2}_b - \rho^{*2}_{12}}
\Rb^{\gamma}
\eeq

\begin{figure}[h]
\centerline{\epsfig{file=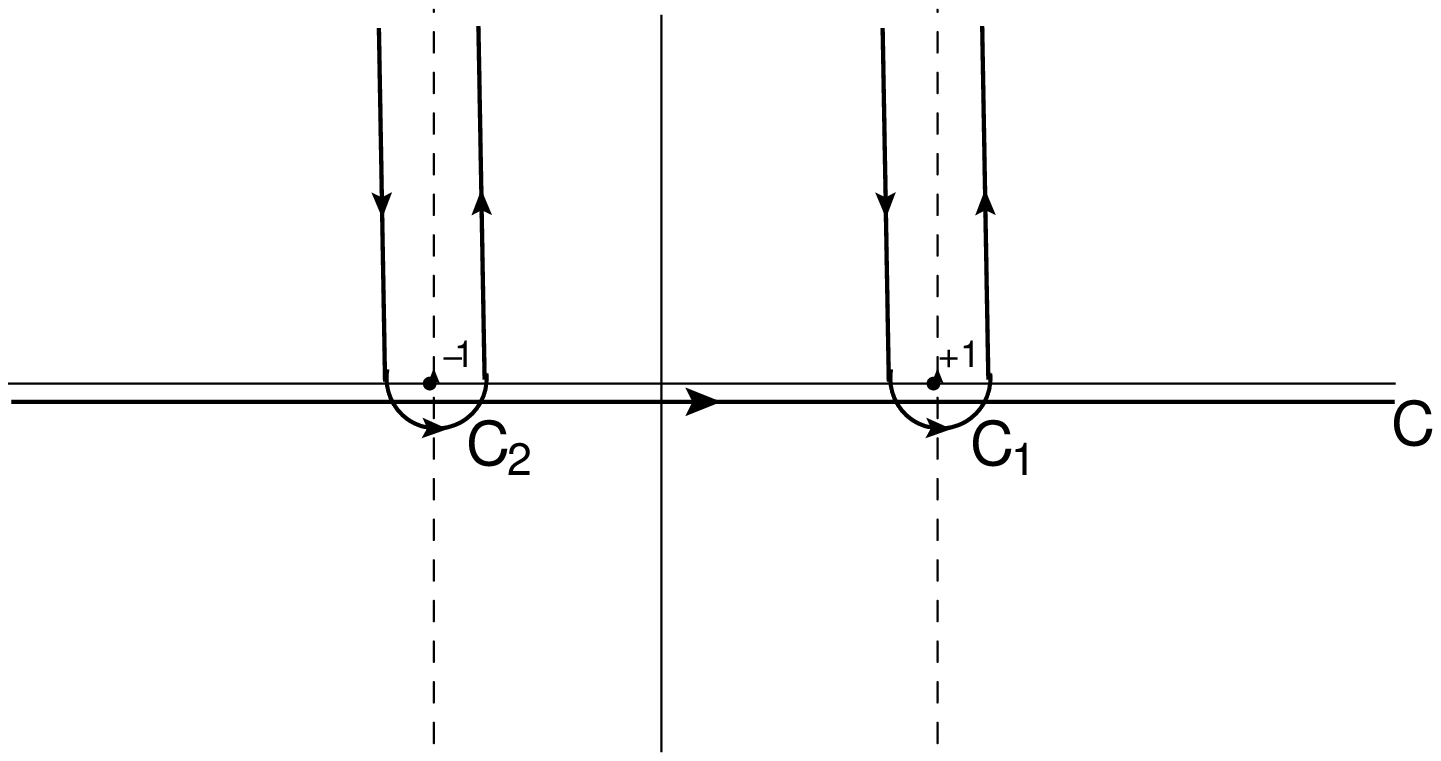,width=60mm}}
\caption{ Contours of  integration in \protect\eq{HAN}.}
 \label{cont} 
\end{figure}


Using new variables $t= \rho_b/\rho_{12}$ and $t^{*} =  \rho^{*}_b/\rho_{12}$
 and the integral representation of Hankel
functions (see formulae {\bf 8.422(1,2)} in Ref. \cite{RY})
\beq \label{HAN}
H^{(1,2)}_\nu\Lb z \Rb\,\,=\,\,\frac{\Gamma\Lb \h
 - \nu\Rb}{\pi i \Gamma\Lb \h\Rb} \Lb \h z \Rb^\nu 
\oint_{C_{1,2}}d t e^{i z t}\Lb t^2 - 1\Rb^{\nu - \h}
\eeq
where contours $C_1$ and $C_2$ are shown in \fig{cont}, we obtain 
\beq\label{NPQ1}
N_\pom\Lb r_1,r_2; Y, Q_T \Rb \,\,=\,\,C^2(\gamma) \,r^2_{12}  \,e^{\bas \, \chi\Lb \ga\Rb\,Y}
\,\Lb\frac{ r^2_1 r^2_2}{r^4_{12}}\Rb^{\gamma}\,\Lb Q^2 r^2_{12}\Rb^ { - \h + \gamma} ~
J_{\h - \gamma}\Lb \rho^*_Q \rho_{12}\Rb J_{\h - \gamma}\Lb \rho_Q \rho^{*}_{12}\Rb
\eeq
where  $ 2 J_\nu(z) = H^{(1)}\Lb z\Rb + H^{(2)}_\nu\Lb z \Rb$; $\vec{r}_{12}
 = \h\Lb \vec{r}_1 - \vec{r}_2\Rb$ and 
\beq \label{CGA}
C(\gamma)\,\,=\,\,2^{-\frac{5}{2} + \gamma} \pi 
 \frac{\Gamma\Lb \h \Rb}{\Gamma\Lb  \gamma\Rb  }
\eeq
Two limits will be useful for further presentation:
\bea
N_\pom\Lb r_1,r_2; Y, Q_T \Rb \,\,&\xrightarrow{Q_T \to 0}&\,\,\,
C^2(\gamma)\,r^2_{12}  \,e^{\bas \, \chi\Lb \ga\Rb\,Y}
\,\Lb\frac{ r^2_1 r^2_2}{r^4_{12}}\Rb^{\gamma}\label{NPQ2}\\
&\xrightarrow{Q^2_T\,r^2_{12}\,\,\gg\,\,1 }&\,\,\frac{2}{\pi}\,C^2(\gamma)
 \,r^2_{12}  \,e^{\bas \, \chi\Lb \ga\Rb\,Y}
\,\Lb\frac{ r^2_1 r^2_2}{r^4_{12}}\Rb^{\gamma}\,\Lb Q^2 r^2_{12}\Rb^ {
 - 1 + \gamma} \cos^2\Lb \pi \gamma/2\Rb e^{i \vec{Q}
 \cdot \vec{r}_{12}}\label{NPQ3}
\eea

\eq{NPQ2} can be re-written at $r_1 \ll r_2$ in the form :
\bea \label{NPQ4}
N_\pom\Lb r_1,r_2; Y, Q_T \Rb \,\,&\xrightarrow{Q_T \to 0,
 r_1 \,\ll\,r_2}&\,\,\,C^2(\gamma)\,r^2_{12} \Lb r^2_1 Q^2_s\Lb Y, r_2\Rb \Rb^{ \gamma} ~~~~\mbox{with}~~Q^2_s\,\,=\,\,\frac{1}{r^2_2}\,e^{\bas \,\frac{ \chi\Lb \ga\Rb}{\gamma}\,Y};\nn\\
& \xrightarrow{Q_T  r_2\,\gg\,1, r_1 \,\ll\,r_2}&\,\,\frac{2}{\pi}\,
C^2(\gamma) \,\cos^2\Lb \pi \gamma/2\Rb e^{i \vec{Q} \cdot \vec{r}_{12}} 
 \,e^{\bas \, \chi\Lb \ga\Rb\,Y}\frac{1}{Q^2_T}\Lb Q^2_T \,r^2_1\Rb ^{\gamma}
\eea

    \subsubsection{General formula}
    In this subsection we calculate the cross section for the
 inclusive production of a gluon jet with  transverse momentum
 $p_\bot$ at rapidity $Y_1$, in the collision of two dipoles with
  sizes $r_1$ and $r_2$, at rapidity $Y$, and at impact parameter $b$.
  The general formula which shows  $k_T$-factorization \cite{KTF}, has
 been derived in Ref.\cite{KTINC} and has the form
    \bea \label{INC}
\frac{d \sigma}{d^2 b\, d Y_1 \,d^2 p_{\bot}}&= & \\
& & \frac{2C_F}{\alpha_s (2\pi)^4}\,\frac{1}{p^2_\bot}\int \!\!d^2 \vec
 B \,d^2 \vec r_\bot\,e^{i \vec{p}_\bot\cdot \vec{r}_\bot}\,\,\nabla^2_\bot\,
N^G\Lb \h Y - y_1; r_\bot,r_{1}; b \Rb\,\,\nabla^2_\bot\,N^G\Lb  \h Y +  y_1;
 r_\bot, r_2; |\vec b-\vec B| \Rb\nn
\eea
  where 
  \beq \label{INC1}
N^G\Lb Y; r_\bot,r_i; b \Rb\,\,=\,\,2 \,N\Lb Y; r_\bot , r_i ; b \Rb\,\,-\,\,N^2\Lb Y; r_\bot, r_i; b \Rb,
\eeq  
  For one Pomeron exchange \eq{INC1} reduces to the following equation
   \beq \label{INC2}
   N^G_\pom \Lb Y; r_\bot,r_i; b \Rb\,\,=\,\,2 \,N_\pom\Lb Y; r_\bot , r_i
 ; b \Rb   
   \eeq
  Plugging \eq{INC2} into \eq{INC} we have
  \bea \label{INC3}
 &&  N^{\mbox{\tiny incl}}_\pom\Lb Y, r_1, r_2, b, p_{\bot}, Y_1\Rb = \\
   && \frac{8\,C_F}{\alpha_s (2\pi)^4}\,\frac{1}{p^2_\bot}\int \,d^2 \vec B
 \,d^2 \vec r_\bot\,e^{i \vec{p}_\bot\cdot \vec{r}_\bot}\,\,\nabla^2_\bot\,N_\pom\Lb Y_1; r_\bot,r_{1}; B \Rb\,\,\nabla^2_\bot\,N_\pom\Lb  Y - Y_1; r_\bot, r_2; |\vec b-\vec B| \Rb\nn
 \eea 
 Note, that $b$ is the difference of the
 impact parameters between scattering dipoles, while $B$ is the
 impact parameter of the produced gluon with respect to  the dipole
 of size $r_1$.

 It  is more convenient to use
 $ \nabla^2_\bot\,N_\pom\Lb Y_1; r_\bot,r_{1}; b \Rb$ in
 momentum representation, namely,
\beq \label{INC5}
\int d^2 b\, e^{i \vec{Q}_T \cdot \vec{b}}~\nabla^2_\bot
\,N_\pom\Lb Y_1; r_\bot,r_{1}; b \Rb \,=\,\nabla^2_\bot\,N_\pom\Lb Y_1;
 r_\bot,r_{1}; Q_T \Rb  
\eeq

Using \eq{NPQ1} for $N_\pom\Lb Y_1; r_\bot,r_{1}; Q_T \Rb  $
 and $\nabla^2_r\,\,=\,\,4\, \partial_\rho \partial_{\rho^*}$
 we obtain( denoting $r_\bot \equiv r_0$)
\bea \label{INC6}
\nabla^2_\bot\,N_\pom\Lb Y_1; r_\bot,r_{1}; Q_T \Rb\,&=&\,\int^{\epsilon
 + i \infty}_{\epsilon - i \infty} \frac{d \gamma}{2 \pi i}\,\,4\,\,
C^2(\gamma) \,r^2_{01}  \,e^{\bas \, \chi\Lb \ga\Rb\,Y}
\,\Lb\frac{ r^2_1 r^2_0}{r^4_{01}}\Rb^{\gamma}\,\Lb Q^2 r^2_{01}\Rb^ {
 - \h + \gamma} \\
&\times& \Bigg\{ \Big(\frac{\gamma}{\rho_r} - \frac{\h + \gamma}{\rho_{01}}
 \Big)\,J_{\h - \gamma}\Lb\rho^*_Q \rho_{01}\Rb + \h\rho^*_Q\Lb J_{-\h -
 \gamma}\Lb \rho^*_Q \rho_{01}\Rb\,-\,J_{-\frac{3}{2} - \gamma}\Lb \rho^*_Q
 \rho_{01}\Rb\Rb\Bigg\}\nn\\
&\times& \Bigg\{\Big( \frac{\gamma}{\rho^*_r} - \frac{\h +
 \gamma}{\rho^*_{01}} \Big)\, J_{\h - \gamma}\Lb\rho_Q \rho^*_{01}\Rb +  \h\rho_Q\Lb J_{-\h - \gamma}\Lb \rho_Q \rho^*_{01}\Rb\,-\,J_{-\frac{3}{2} - \gamma}\Lb \rho_Q \rho^*_{01}\Rb\Rb\Bigg\}\nn
\eea   

We need  to estimate
\beq \label{INC7}
N^{\mbox{\tiny incl}}_\pom\Lb Y, r_1, r_2, Q_T, p_{\bot},
 Y_1\Rb \,=\,\int d^2 b ~e^{i \vec{Q}_T \cdot
 \vec{b}}\,N^{\mbox{\tiny incl}}_\pom\Lb Y, r_1, r_2, b, p_{\bot}, Y_1\Rb
\eeq
 for calculating  the diagrams of \fig{nimcor2sh}-a.  From
 \eq{INC3} and \eq{INC6} we obtain
 \bea\label{INC8}
&& N^{\mbox{\tiny incl}}_\pom\Lb Y, y_1, r_1, r_2, Q_T, p_{\bot}, Y_1\Rb \,=\\
  &&~~~~~~~~~~~~~~~~~ \, \frac{8\,C_F}{\alpha_s (2\pi)^4}\,
\frac{1}{p^2_\bot}\int d^2 \vec r_\bot\,e^{i \vec{p}_\bot\cdot
 \vec{r}_\bot}\,\,\nabla^2_\bot\,N_\pom\Lb y_1; r_\bot,r_{1}; Q_T
 \Rb\,\,\nabla^2_\bot\,N_\pom\Lb  Y - y_1; r_\bot, r_2; Q_T \Rb\nn~
 \eea
   
   Note,  the dependence on $y_1$  is very weak 
 since $N_\pom \Lb y_1; r_\bot,r_{1}; Q_T \Rb   \,\propto \exp\Lb
 \bas \chi(\gamma) \,y_1\Rb$.
     
 \subsubsection{Azimuthal angle  dependance}
  As we have discussed in the introduction, the azimuthal 
angle correlation arises from the terms $\Lb \vec{p}_{\bot,1}
 \cdot \vec{Q}_T\Rb^2$ and $ \Lb \vec{p}_{\bot,2} \cdot \vec{Q}_T\Rb^2$,
 after integration over $Q_T$ in the Pomeron loop in the diagram of
 \fig{nimcor2sh}-a,  since \\$\int d^2 Q_T    \Lb \vec{p}_{\bot,1}
 \cdot \vec{Q}_T\Rb^2  \Lb \vec{p}_{\bot,2} \cdot \vec{Q}_T\Rb^2
 \to \Lb  \vec{p}_{\bot,1} \cdot      \vec{p}_{\bot,2} \Rb^2$.  
  Such terms in the coordinate representation that we are using
 here, stem from the terms  $\Lb \vec{r}_{12} \cdot \vec{Q}_T\Rb^2$
 and $ \Lb \vec{r}^{\,'}_{12} \cdot \vec{Q}_T\Rb^2$ in $N_\pom\Lb
 r_1, r_2, Y, Q_T\Rb$ and $N_\pom\Lb r'_1, r'_2, Y, Q_T\Rb$  
 (see \eq{NPQ1}). These terms come from $J_{\h - {\gamma}}\Lb
 \rho^{*}_Q\,\rho_{12}\Rb \,J_{\h - {\gamma}}\Lb \rho_Q\,\rho^{*}_{12}\Rb$.
 For small $Q_T$ we can see how these terms appear by expanding $J_{\h -
 {\gamma}}$.

   Indeed,
   \bea \label{AC1}
   &&\Lb Q^2 r^2_{12}\Rb^ {- \h + \gamma} \,J_{\h - {\gamma}}\Lb \rho^*_Q\,\rho_{12}\Rb \,J_{\h -
 {\gamma}}\Lb \rho_Q\,\rho^*_{12}\Rb = \\
 &&=\,\,  \Lb \frac{1}{2^{\h - {\gamma}}\,\Gamma\Lb \frac{3}{2} -
 {\gamma}\Rb}\Rb^2 \Big\{ 1 + \frac{1}{2 ( - 3 + 2 \,{\gamma})}
   Q^2_T  r^2_{12} e^{ 2 i ( \phi - \psi)}  \Big\} \Big\{1 + \frac{1}{2
 ( - 3 + 2 \,{\gamma})}
   Q^2_T r^2_{12} e^{ -2 i ( \phi - \psi)}  \Big\}\nn\\
   &&\to\,\, \Lb \frac{1}{2^{\h - {\gamma}}\,\Gamma\Lb \frac{3}{2}
 - {\gamma}\Rb}\Rb^2   \Big\{ 1 +\frac{1}{ ( - 3 + 2 \,{\gamma})}
 Q^2_T r^2_{12} \cos\Lb 2  ( \phi - \psi)\Rb + \Lb \frac{1}{ 2 (
 - 3 + 2 \,{\gamma})}\Rb^2 Q^4_T r^4_{12}\Big\}\nn
   \eea
   \bea
     &&\to\,\,  \Lb \frac{1}{2^{\h - {\gamma}}\,\Gamma\Lb \frac{3}{2}
 - {\gamma}\Rb}\Rb^2   \Big\{ 1 +\frac{1}{ ( - 3 + 2 \bar{\gamma})}\Lb 
 2 \Lb \vec{Q}_T \cdot \vec{r}_{12}\Rb^2  - Q^2_T r^2_{12}\Rb  
 + \Lb \frac{1}{ 2 ( - 3 + 2 \,{\gamma})}\Rb^2 Q^4_T r^4_{12}\Big\} \nn
     \eea  
 In \eq{AC1} we use the representation of  complex numbers in the
 polar coordinates, for example, $\rho_Q = Q e^{i
 \phi}$ and $\rho^{*}_Q = Q e^{-i \phi}$. The same
 type of contributions come from \eq{INC6}.

  For $Q_T r_{12} \gg 1$ we have the same features since
  
  \bea
&&N_\pom\Lb r_1,r_2; Y, Q_T \Rb \,\,\xrightarrow{Q^2_T\,r^2_{12}\,\,\gg\,
\,1 }\label{AC2}
\\
&&~,\int^{\epsilon + i \infty}_{\epsilon - i \infty}
 \frac{d \gamma}{2 \pi i}\,\,\,\,C^2({\gamma}) \,\frac{2}{\pi}\,
r^2_{12}  \,e^{\bas \, \chi\Lb \ga_{cr}\Rb\,Y}
\,\Lb\frac{ r^2_1 r^2_2}{r^4_{12}}\Rb^{{\gamma}}\,\Lb Q^2_T r^2_{12}\Rb^
 { - 1 + {\gamma}} \cos^2\Lb \pi {\gamma}/2\Rb e^{i \vec{Q}_T \cdot \vec{r}_{12}}\nn\\
&& \Big\{ (1/8)({\gamma}\,({\gamma} - 2) ( 1 - {\gamma}^2))
  + Q^2_T r^2_{12}\,e^{i 2 (\phi - \psi)}\Big\} \Big\{(1/8)(\bar{\gamma}\,
({\gamma} - 2) ( 1 - {\gamma}^2)) + Q^2_T r^2_{12}\,e^{- i
 2 (\phi - \psi)}\Big\}\Big{/}Q^4_T r^4_{12}\nn\\
&& =\,\,\,\int^{\epsilon + i \infty}_{\epsilon - i \infty}
 \frac{d \gamma}{2 \pi i}\,\,\,C^2({\gamma}) \, \,\frac{2}{
 \pi}\,r^2_{12}  \,e^{\bas \, \chi\Lb \ga_{cr}\Rb\,Y}
\,\Lb\frac{ r^2_1 r^2_2}{r^4_{12}}\Rb^{{\gamma}}\,\Lb Q^2_T r^2_{12}\Rb^
 { - 3 + {\gamma}} \cos^2\Lb \pi {\gamma}/2\Rb e^{i \vec{Q}_T \cdot
 \vec{r}_{12}} \nn\\
&&\Bigg\{ \Big[(1/8)({\gamma}\,(\bar{\gamma} - 2)
 ( 1 - {\gamma}^2)) \Big]^2 +  (1/4)({\gamma}\,({\gamma} - 2) ( 1 - {\gamma}^2))  \Lb 2 \Lb \vec{Q}_T \cdot \vec{r}_{12}\Rb^2 - Q^2_T r^2_{12}\Rb + Q_T^4 r^4_{12}\Bigg\}\nn\eea

However, the largest contribution  at $r \ll r_1 $ and $r
 \ll r_2$  comes from \eq{INC6}, which can be re-written as

\bea \label{AC3}
&&\nabla^2_\bot\,N_\pom\Lb Y_1; r_\bot,r_{1}; Q_T \Rb\,\,\to \\
 &&~~~~~~~~~\,\,\int^{\epsilon + i \infty}_{\epsilon - i \infty}
 \frac{d \gamma}{2 \pi i}\,\,\,\,C^2({\gamma}) \,r^2_{01} 
 \,e^{\bas \, \chi\Lb \ga\Rb\,Y}
\,\Lb\frac{ r^2_0 r^2_1}{r^4_{01}}\Rb^{{\gamma}}\,\Lb Q^2
 r^2_{01}\Rb^ { - \h + {\gamma}}\frac{{\gamma}^2}{ r^2}\,
 J_{\h - {\gamma}}\Lb \rho^*_Q \rho_{01}\Rb \, J_{\h -
 {\gamma}}\Lb \rho_Q \rho^*_{01}\Rb\nn
\eea
where $r_0 \equiv r_\bot$.

Note, that at $Q_T \to 0$ \eq{AC3} reduces to
\bea\label{AC31}
&&\nabla^2_\bot\,N_\pom\Lb Y_1; r_\bot,r_{1}; Q_T \Rb\,\,\to\\
&& \,\,\int^{\epsilon + i \infty}_{\epsilon - i \infty} \frac{d \gamma}{2 \pi i}\,\,4\,C^2({\gamma}) \,r^2_{01}  \,e^{\bas \, \chi\Lb \ga\Rb\,Y}
\,\Lb\frac{ r^2_0 r^2_1}{r^4_{01}}\Rb^{{\gamma}}\,\Lb \frac{2^{-\h {\gamma}}}{\Gamma\Lb 3/2 - {\gamma}\Rb}\Rb^2\frac{{\gamma}^2}{ r^2}\Big\{ 1 - \frac{1}{( 3 - 2 {\gamma})}\Lb \rho^2_Q \rho^{*2}_{01}\,\,+\,\, \rho^{*2}_Q \rho^{2}_{01}\Rb\Big\} \nn\\
&&= \,\,\int^{\epsilon + i \infty}_{\epsilon - i \infty} \frac{d \gamma}{2 \pi i}\,\,4\,C^2({\gamma}) \,r^2_{01}  \,e^{\bas \, \chi\Lb \ga\Rb\,Y}
\,\Lb\frac{ r^2_0 r^2_1}{r^4_{01}}\Rb^{{\gamma}}\,\Lb \frac{2^{-\h {\gamma}}}{\Gamma\Lb 3/2 - {\gamma}\Rb}\Rb^2\frac{{\gamma}^2}{ r^2}\Bigg\{ 1 - \frac{2}{ ( 3 - 2 { \gamma})}\, \Lb \vec{Q}_T \vec{r}_{01} \Rb^2 \Bigg\}\nn\\
&&= \,\,\int^{\epsilon + i \infty}_{\epsilon - i \infty} \frac{d \gamma}{2 \pi i}\,\,4\,\,C^2({\gamma}) \,r^2_{01}  \,e^{\bas \, \chi\Lb \ga\Rb\,Y}
\,\Lb\frac{ r^2_0 r^2_1}{r^4_{01}}\Rb^{{\gamma}}\,\Lb \frac{2^{-\h {\gamma}}}{\Gamma\Lb 3/2 - {\gamma}\Rb} \Rb^2\frac{{\gamma}^2}{ r^2}\Bigg\{ 1 - \frac{2}{ ( 3 - 2 \,{ \gamma})}\,\Lb \vec{Q}_T \vec{r} \Rb^2\Bigg\}\nn \eea  

To calculate the production of the gluon with the transverse momentum
 $p_{1,\bot}$,
we need to plug  \eq{AC31} into  \eq{FINSOL} and \eq{INC8}.

 \subsection{Angular dependence of the double  inclusive cross section}
The contribution to the double inclusive production from the
 diagram of \fig{nimcor2sh}  takes the general form
\bea \label{ACFD1}
&&\frac{d^2 \sigma^{\mbox{\tiny 2 parton showers}}}{ d
 y_1\,\,d y_2,d^2 p_{1,\bot} d^2 p_{2,\bot}} \,\,=\\
&& \,\,\sum^{i = 2,j=2}_{i=1,j=1} \alpha^2_i \,\alpha^2_j
 \int g^2_i\Lb Q_T\Rb\,g^2_j\Lb Q_T\Rb\, N^{\mbox{\tiny incl}}_\pom\Lb Y,y_1,
 r_1, r_2, Q_T, p_{1,\bot}, Y_1\Rb  \,  N^{\mbox{\tiny incl}}_\pom\Lb Y, y_2,
 r_1, r_2, Q_T, p_{2, \bot}, Y_1\Rb \,\frac{d^2 Q_T}{4 \pi^2}\nn
\eea
  where $\alpha_1 = \alpha, \alpha_2= \beta$ and
 $g_i\Lb Q_T\Rb\,\,=\,\,g_i \int d^2 b \,e^{i \vec{Q}_T
 \cdot\vec{b}}\,S_\pom\Lb b, m_i\Rb\,=\,g_i/(1 + Q^2_T/m^2_i)^2$,
 as it follows from \eq{SB}.  
  
  Substituting \eq{AC31} into \eq{INC8} we obtain 
  \bea \label{ACFD2}
  && N^{\mbox{\tiny incl}}_\pom\Lb Y,y_1, r_1, r_2, Q_T, p_{1\bot},
 Y_1\Rb\,\,=\,\,\frac{8\,C_F}{\bas (2\pi)^4}\,\frac{1}{p^2_{1\bot}}
\int d^2 \vec r_0\,e^{i \vec{p}_{1,\bot}\cdot \,\vec{r}_0}\,\\
     &&\int^{\epsilon + i \infty}_{\epsilon - i \infty} \frac{d
 \gamma_1}{2 \pi i}\,\,4\,\,C^2({\gamma_1}) \,r^2_{01}  \,e^{\bas \,
 \chi\Lb \ga_1\Rb\,\Lb \h Y - y_1\Rb}
\,\Lb\frac{ r^2_0 r^2_1}{r^4_{01}}\Rb^{{\gamma_1}}\,\Lb \frac{2^{-\h
 {\gamma_1}}}{\Gamma\Lb 3/2 - {\gamma_1}\Rb} \Rb^2\frac{\gamma_1^2}{
 r^2_0}\Bigg\{ 1 - \frac{2}{ ( 3 - 2 \,{ \gamma_1})}\,\Lb \vec{Q}_T
 \vec{r}_0 \Rb^2\Bigg\}\nn\\
&& \int^{\epsilon + i \infty}_{\epsilon - i \infty} \frac{d
 \gamma_2}{2 \pi i}\,\,4\,\,C^2({\gamma_2}) \,r^2_{01}  \,e^{\bas
 \, \chi\Lb \ga_2\Rb\,\Lb \h Y + y_1\Rb}
\,\Lb\frac{ r^2_0 r^2_2}{r^4_{02}}\Rb^{{\gamma_2}}\,\Lb
 \frac{2^{-\h {\gamma_2}}}{\Gamma\Lb 3/2 - {\gamma_2}\Rb} \Rb^2
\frac{\gamma_2^2}{ r^2_0}\Bigg\{ 1 - \frac{2}{ ( 3 - 2
 \,{ \gamma_2})}\,\Lb \vec{Q}_T \vec{r}_0 \Rb^2\Bigg\}
\nn \eea     
      
   Integrating first over $d^2 r_0  $ we obtain
     \bea \label{ACFD3}
  && N^{\mbox{\tiny incl}}_\pom\Lb Y,y_1, r_1, r_2, Q_T, p_{1\bot},
 Y_1\Rb\,\,=\,\,\frac{128\,C_F}{\bas (2\pi)^4}\,r^2_1 r^2_2
       \int^{\epsilon + i \infty}_{\epsilon - i \infty} \frac{d
 \gamma_1}{2 \pi i}\, \int^{\epsilon + i \infty}_{\epsilon - i
 \infty} \frac{d \gamma_2}{2 \pi i}\,\,C^2({\gamma_1})
 \,\,C^2({\gamma_2})\\
       &&  \,e^{\bas \,\Lb  \chi\Lb \ga_1\Rb\,\Lb \h Y -
 y_1\Rb \,+\,\chi\Lb \ga_2\Rb\,\Lb \h Y +y_1\Rb\Rb}
\,\Lb\frac{1}{r^2_{1}\,p^2_{1\bot}}\Rb^{{\gamma_1}}\,
\Lb\frac{1}{r^2_{2}\,p^2_{1\bot}}\Rb^{{\gamma_2}}\,\Lb
 \frac{2^{- {\gamma_{12}}}}{\Gamma\Lb 3/2 - {\gamma_1}\Rb^2\Gamma\Lb 3/2
 - {\gamma_1}\Rb^2} \Rb^2\,\frac{\Gamma\Lb - 1 +
 \gamma_{12}\Rb}{\Gamma\Lb 2 - \gamma_{12}\Rb}\nn\\
&&\Bigg\{ 1 +   4\,\Lb\gamma_{12}-1\Rb \gamma_{12}\Lb
 \frac{2}{ ( 3 - 2 \,{ \gamma_1})} \, \,+\,\,\frac{2}{ ( 3
 - 2 \,{ \gamma_2})}\Rb \Lb\frac{ \vec{Q}_T \cdot
 \vec{p}_{1,\bot}}{p^2_{1\bot}} \Rb^2\Bigg\}\nn
\eea        
   In \eq{ACFD3} we denote $\gamma_{12} = \gamma_1 + \gamma_2$,
 consider $r_0 \ll r_1(r_2)$, and neglected the contributions of
 the order of $Q^4_T$.
   
 For further estimates, we need to   return to a general formula of 
\eq{INC}. 
 We know that as a result of the shadowing corrections
 $N^G\Lb Y; r_\bot,r_{1}; b \Rb\,\,\to\,\,1 $ for large values of $Y$. It
 means that 
 $ \nabla^2_\bot\,N^G\Lb Y; r_\bot,r_{1}; b \Rb \,\,\to\,\,0$ in the
 saturation region, where $ r^2_\bot Q^2_s\Lb Y\Rb \,\gg\,1$. 
 Such  behavior stems from the diagrams of \fig{nimcor2sh}-b, but not
 from the first diagram that we are presently considering.  
 Consequently,
  we have $ \nabla^2_\bot\,N^G\Lb Y; r_\bot,r_{1}; b \Rb
 \,\,\to\,\,0 $ which vanishes both at $r^2_\bot Q^2_s\Lb Y\Rb \,\gg\,1$
 and at  
     $ r^2_\bot Q^2_s\Lb Y\Rb \,\ll\,1$, and the main contribution 
 originates for  the value of $r$ in the vicinity of the saturation
 scale $ r^2_\bot Q^2_s\Lb Y\Rb \,\,\approx\,\,1  $.
  As we have discussed,
  in the vicinity of the saturation scale $\gamma_1 = \gamma_2 = \bar
 \gamma =1 - \gamma_{cr} = 0.63$ (see \eq{GACR} and \eq{GSNB1}).
  The second observation which simplifies the estimates, is that in
 our approach $r_1 = r_2 \, \sim \,1/m  $ and $m \gg m_1$ and $m_2$
 (see Table 1).
  Since the typical $Q_T$ in the integration is approximately $m_1$ or 
$m_2$,
 we can neglect the $Q_T$ dependance of  the BFKL Pomeron .
  
  Therefore, we can write the double inclusive production cross section
 in the following form
      
\beq \label{ACFD4}
\frac{d^2 \sigma^{\mbox{\tiny 2 parton showers}}}{ d
 y_1\,\,d y_2,d^2 p_{1\bot} d^2 p_{2\bot}} \Bigg{/} \int \frac{d
 \phi}{2 \pi} \frac{d^2 \sigma^{\mbox{\tiny 2 parton showers}}}{ d
 y_1\,\,d y_2,d^2 p_{1\bot} d^2 p_{2\bot}}\,\,=\,\, 1 \,
\,+\,\frac{\kappa}{p^2_{1\bot}\,p^2_{2\bot}}\cos\Lb 2 \phi\Rb
 \eeq
     where $\kappa$ is equal to
     \beq \label{KAPP}
     \kappa \,=\,\Big( \frac{8 \bar \gamma\Lb 2 \bar \gamma -1\Rb}{
 3 - 2 \,\bar{ \gamma}}\Big)^2\,\,
      \frac{ \sum^{i = 2,j=2}_{i=1,j=1} \alpha^2_i \,\alpha^2_j
 \,g^2_i\,g^2_j   \int d Q^2_T \,   \frac{Q^4_T}{\Lb 1 +
 Q^2_T/m^2_i\Rb^2\,\Lb 1 + Q^2_T/m^2_j\Rb^2  }}{ \sum^{i =
 2,j=2}_{i=1,j=1} \alpha^2_i \,\alpha^2_j \,g^2_i\,g^2_j  
 \int d Q^2_T \,   \frac{1}{\Lb 1 + Q^2_T/m^2_i\Rb^2\,\Lb 1
 + Q^2_T/m^2_j\Rb^2 }}
     \eeq
     
\begin{figure}[h]
\begin{center}
\epsfig{file=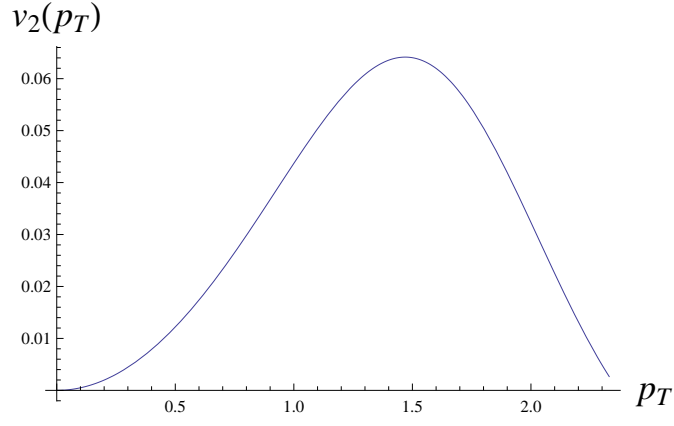,width=90mm}
\end{center}
\caption{$\nu_{22} \, \equiv \,v_{22} $ versus $p_T \equiv p_\bot $ for 
$p_{1\bot}=p_{2\bot}=p_T$.
 $R = 3/Q_s$ at $Q_s = 1\,GeV$ .}
 \label{v2} 
\end{figure}

     In our model $i = j =1$ gives the largest contribution, due to large
 value of $g_1$(see Table 1), and we obtain $\kappa = 0.04\,GeV^4$.
     The contribution of the term proportional to $\cos\Lb 2 \phi\Rb$
 depends on the value of $p_\bot$. Actually, we can trust \eq{ACFD4}
 for $p_{i,\bot} \,\geq\,Q_s(Y)$. Integrating over $p_{1\bot}$ and
 $p_{2\bot}$ we expect that the contribution to the correlation
 function will be 
     equal to
     
     \beq \label{ACFDCOR}
     R\Lb y_1,y_2,\phi\Rb\,=\,R\Lb y_1,y_2\Rb \Big(\frac{\kappa}
{Q^4_s\Lb Y\Rb}\Big)
 \,\cos\Lb 2 \phi\Rb\,=\,2 \,v^2_2\,\cos\Lb 2 \phi\Rb   
  \eeq
     
     leading to $R\Lb \h Y,\h Y,\phi\Rb = 2\,0.06\cos\Lb 2 \phi\Rb $ for
 $Q_s \approx 1\,GeV$ or $v_2 = 0.23$.  This value is in a good agreement
 with the estimates for this correlation from the elliptic flow\cite{V2PP}
 and experimental data\cite{CMSACOR,ATLASACOR}.   \eq{ACFD4} leads to $v_2 =
 R\Lb y_1,y_2\Rb \kappa/p^2_{1,\bot}\,p^2_{2,\bot}$. However, we can
 trust this      $p_\bot$ dependence only for $p^2_{1\bot} > Q_s$ 
 and $   p^2_{2\bot} > Q_s$ . We should introduce the shadowing
 corrections to reproduce the behaviour
     of $v_2$ for $p^2_{1\bot} < Q_s$  and $   p^2_{2\bot} <Q_s0$. We 
will do this in the next section for our model,  but here we estimate
 the influence of the shadowing correction by integrating \eq{ACFD2}
 over $r_0$ in the limits $ 0 \,<\,r_0 \,<\,R \sim 1/Q_s$. Indeed, as
 has been mentioned,  $N_G \to 0$ for $r_0 \gg 1/Q_s$. In \fig{v2} we
 plot the $v_2$ dependence for $p_{1\bot}=p_{2\bot}$ for $Q_s = 
1\,GeV$
 and choosing $R = 3/Q_s$. One can see that $v_2$ decreases at $p_\bot
 < Q_s$.
     
 \section{Formula for azimuthal correlations of two particles produced
 in two parton showers }
  
  Based on our  experience from calculating  the first diagrams, we
 can estimate the two particle angular correlations that stem for the
 general diagram of \fig{nimcor2sh}-b.  The general formulae have the
 same form as \eq{INC}, which we  used in estimating the contribution
 of the first diagram. However, we now need to calculate $N_G$ in \eq{INC}
 using our model described in section 2.
  As we have seen  for angular correlations, it is essential to use the 
$r$ 
and
 $b$ dependence of the BFKL contribution. Bearing this in mind, we have to 
 generalize \eq{T} replacing it by the following formula
  \beq \label{TT}
 T\Lb r_\bot , b, Y\Rb\,\rightarrow\,T_W\Lb r_\bot , b, Y\Rb\,\,=\,\,
 \phi_0\, \Lb w\,w^*\,Q^2_s\Lb Y, b\Rb\Rb^{\bar \gamma}
  \eeq
  where $w\,w^*$ is given by \eq{W} with the arguments $r_2 = r_\bot =
 r_0$ and $r_2 = 1/m$, and $Q_s$ is given by \eq{QS}. In \eq{TT} we
 replace $r^2_\bot $ in \eq{T} by $w\,w^*$, since $T_W$ describes the
 behavior of the scattering amplitude in the vicinity of the saturation scale.

  The expression of $N_G$ is the direct generalization of \eq{IN} and
 has the following form
  \beq \label{NGAC}
  N^i_G\Lb r_\bot ,r_1; Y, Q_T \Rb\,\,\,=\,\,\int d^2 b \,\,e^{i 
\vec{Q}_T \cdot \vec{b}}\,\,N^{\mbox{\tiny BK}}\Bigg(\int d^2 b'
 \,G^{\mbox{\tiny  dressed}}\Big( T_W\Lb r_\bot, b',Y\Rb\Big)
 g_i\Lb \vec{b} - \vec{b}^{\,'}\Rb\Bigg)
  \eeq
  Calculating $\nabla^2_{r_\bot}  N^i_G\Lb r_\bot ,r_1; Y, Q_T \Rb$ we
 see that we have two contributions
  
  \bea \label{NGAC1}
&&    \nabla^2_{r_\bot}  N^i_G\Lb r_\bot ,r_1; Y, Q_T
 \Rb\,\,=\,\,\int d^2 b \,d^2 b'\,\,\,e^{i \vec{Q}_T \cdot \vec{b}}\\
&&\Bigg\{
    \nabla^2_{r_\bot}\,G^{\mbox{\tiny  dressed}}\Big(
 T_W\Lb r_\bot, b',Y\Rb\Big)   \frac{d  }{d G^{\mbox{\tiny 
 dressed}} } \,\,+\,\,   \vec{\nabla}_{r_\bot}\,G^{\mbox{\tiny 
 dressed}}\Big( T_W\Lb r_\bot, b',Y\Rb\Big) \cdot \vec{\nabla}_{r_\bot}
\,G^{\mbox{\tiny  dressed}} \Big( T_W\Lb r_\bot, b',Y\Rb\Big)\frac{d^2 
 }{\Lb d G^{\mbox{\tiny  dressed}}  \Rb^2}     \Bigg\}  \nn\\
    &&  N^{\mbox{\tiny BK}}\Bigg(\int d^2 b' \,G^{\mbox{\tiny 
 dressed}}\Big( T_W\Lb r_\bot, b',Y\Rb\Big) g_i\Lb \vec{b} -
 \vec{b}^{\,'}\Rb\Bigg)   \nn   
    \eea
    
    The first term in \eq{NGAC1} is proportional to
 $ \Lb r^2_\bot \Rb^{\bar \gamma - 2}$, while the second
 one is $\propto \Lb r^2_\bot \Rb^{2 \bar \gamma - 2}$.
 Therefore, for small $r_\bot$ we can neglect the second term.

We have that
\bea \label{NGAC2}
&& \frac{d  }{d G^{\mbox{\tiny  dressed}} } N^{\mbox{\tiny BK}}\Bigg(\int
 d^2 b' \,G^{\mbox{\tiny  dressed}}\Big( T_W\Lb r_\bot,
 b',Y\Rb\Big) g_i\Lb \vec{b} - \vec{b}^{\,'}\Rb\Bigg)\,\,=\\
&&~~~~~~~~~~~~~~~~~~~~\,\,g_i \Lb \vec{b} - \vec{b}^{\,'}\Rb 
   \frac{d N^{\mbox{\tiny BK}} \Lb {\cal Z}\Rb}{d {\cal Z}}\Bigg(
 {\cal Z}= \int d^2 b' \,G^{\mbox{\tiny  dressed}}\Big( T_W\Lb
 r_\bot, b',Y\Rb\Big) g_i\Lb \vec{b} - \vec{b}^{\,'}\Rb\Bigg) \nn
 \eea
 with
 \beq \label{NGAC3}
 N'^{\mbox{\tiny BK}} \Lb {\cal Z}\Rb\,\equiv\,
 \frac{d N^{\mbox{\tiny BK}} \Lb {\cal Z}\Rb}{d
 {\cal Z}}\,\,=\,\,  a\,e^{-{\cal Z}}\,+\,\frac{ 1 -
 a}{\Lb 1 \,+\,{\cal Z}\Rb^2}
  \eeq
  
  For small $r_\bot$ 
  \bea   \label{NGAC4}     
  \nabla^2_{r_\bot}\,G^{\mbox{\tiny  dressed}}\Big( T_W\Lb r_\bot,
 b',Y\Rb\Big) & \,\,=\,\,&  \nabla^2_{r_\bot} T_W\Lb r_\bot, b',Y\Rb
 \,\frac{d}{d T}G^{\mbox{\tiny  dressed}}\Big( T = T_W\Lb r_\bot,
 b',Y\Rb\Big) \nn\\
  &  \,\,\equiv\,\,&\underbrace{\nabla^2_{r_\bot}\,N_\pom \Lb r_\bot,
 b',Y\Rb}_{ r_{bot} \to 1/Q_s(Y,b)}\,\,\frac{d}{d T}G^{\mbox{\tiny 
 dressed}}\Big( T = T_W\Lb r_\bot, b',Y\Rb\Big) 
    \eea
     with
     \beq \label{NGAC5} 
  G'\Lb T\Rb\,\equiv\,   \frac{d}{d T}G^{\mbox{\tiny
  dressed}}\Lb T\Rb\,\,=\,\,      
     a^2 e^{-T}+\frac{(a-1)^2 e^{1/T} (T+1) \Gamma
 \left(0,\frac{1}{T}\right)}{T^3}+(a-1) \left(\frac{1-a}{T^2}-\frac{2 a}
{(T+1)^2}\right)
     \eeq

         Plugging \eq{NGAC2} - \eq{NGAC5} into \eq{NGAC1}
 and using that $b' \approx 1/m\,\ll\,\, b\,\approx 1/m_i$ we obtain
   \bea\label{NGAC6}
  \nabla^2_{r_\bot}  N^i_G\Lb r_\bot ,r_1; Y, Q_T
 \Rb\,&=&\,\underbrace{\nabla^2_{r_\bot} N^i_G\Lb r_\bot , Y, Q_T \Rb 
 }_{r_\bot \to 1/Q_s}\,\widetilde{G'}\Lb r_\bot,Y\Rb\,N'^{\mbox{\tiny BK}}_i
  \Lb r_\bot,Y, Q_T\Rb \\
  N'^{\mbox{\tiny BK}}_i\Lb r_\bot,Y, Q_T\Rb& =  & \,\, \int d^2  b \,  
 e^{i \vec{Q}_T \cdot \vec{b}}    \,g_i\Lb b\Rb\,{N'}^{\mbox{\tiny BK}}
 \Big( \widetilde{G}\Lb r_\bot,Y\Rb\, g_i\Lb b \Rb\Big)\nn
  \eea
  
  where
  \bea \label{NGAC7}
  \widetilde{G}\Lb r_\bot,Y\Rb &\,\,=\,\,& \int \,d^2 b'\,G\Big(
 T\Lb r_\bot,b',Y\Rb \Big) \nn\\
  \widetilde{G'}\Lb r_\bot,Y\Rb &\,\,=\,\,& \int \,d^2 b'\,S(b',
 m)\,G'\Big( T\Lb r_\bot,b',Y\Rb \Big)
  \eea
     Using \eq{AC31}, in which we substitute $\gamma = \bar \gamma$  we 
obtain 
     \bea \label{NGAC8}
     &&\nabla^2_\bot\,N^i_G\Lb Y; r_\bot; Q_T \Rb\,\,\to \,\,\,4\,
\,C^2({\bar \gamma}) \,r^2_{01}  \,e^{\bas \, \chi\Lb \bar\ga\Rb\,Y}\\
     &&\times\,\,
\,\Lb\frac{ r^2_\bot r^2_1}{r^4_{01}}\Rb^{{\bar \gamma}}\,\Lb \frac{2^{-\h
 {\bar\gamma}}}{\Gamma\Lb 3/2 - {\bar\gamma}\Rb} \Rb^2\frac{{\bar\gamma}^2}
{ r^2}\Bigg\{ 1 - \frac{2}{ ( 3 - 2 \,{\bar \gamma})}\,\Lb \vec{Q}_T 
\vec{r}_\bot \Rb^2\Bigg\}\widetilde{G'}\Lb
 r_\bot,Y\Rb\,N'^{\mbox{\tiny BK}}_i
  \Lb r_\bot,Y, Q_T\Rb\nn \\
  && =\,\,{\cal C}\Lb \bar \gamma\Rb  \,e^{\bas \,
 \chi\Lb \bar\ga\Rb\,Y}\,r^2_{01}\Lb\frac{ r^2_\bot
 r^2_1}{r^4_{01}}\Rb^{{\bar \gamma}}\frac{1}{r^2_\bot} 
 \Bigg\{ 1 - \frac{2}{ ( 3 - 2 \,{\bar \gamma})}\,\Lb
 \vec{Q}_T \vec{r}_\bot \Rb^2\Bigg\}\widetilde{G'}\Lb
 r_\bot,Y\Rb\,N'^{\mbox{\tiny BK}}_i
  \Lb r_\bot,Y, Q_T\Rb\nn  \eea  
      Using \eq{NGAC8} we can re-write the double inclusive cross section
 in the form (see \eq{ACFD4})
      
     \beq \label{NGAC9}
\frac{d^2 \sigma^{\mbox{\tiny 2 parton showers}}}{ d
 y_1\,\,d y_2,d^2 p_{1\bot} d^2 p_{2\bot}} \Bigg{/}
 \int \frac{d \phi}{2 \pi} \frac{d^2 \sigma^{\mbox{\tiny
 2 parton showers}}}{ d
 y_1\,\,d y_2,d^2 p_{1\bot} d^2 p_{2\bot}}\,\,=\,\,
 1 \,\,+\, 2 v_{22}\Lb p_{1T}, p_{2T}\Rb\,\cos\Lb 2 \phi\Rb
 \eeq      
 where
 \beq \label{RCAL}
v_{22}\Lb p_{1T}, p_{2T}\Rb \,\,=\,\,  N\Lb p_{1\bot}, p_{2\bot},
  Y\Rb\Big{/} D\Lb
 p_{1\bot}, p_{2\bot},  Y\Rb
 \eeq
  where 
  \bea \label{RCALN}
 &&N\Lb p_{1\bot}, p_{2\bot},Y, y_1,y_2\Rb\,\,=\,\,\\
 &&\frac{1}{4}\, \sum^{i,j}_{i=1,j=1}\alpha^2_i
 \alpha^2_j \Bigg\{ \int \frac{ d r^2_\bot }{r^2_\bot}J_2\Lb p_{1\bot}
 r_\bot\Rb\Lb\frac{ r^2_\bot}{r^2_{1}}\Rb^{2{\bar \gamma}}
 \frac{4}{ ( 3 - 2 \,{\bar \gamma}) }  \widetilde{G'}\Lb
 r_\bot,\h Y - y_1 \Rb\,\widetilde{G'}\Lb r_\bot,\h Y + y_1 \Rb\Bigg\}\nn\\
 & \times&\,\, \Bigg\{ \int  \frac{d r'^2_\bot}{r'^2_\bot}  J_2\Lb p_{1\bot}
 r'_\bot\Rb\Lb\frac{ r'^2_\bot}{r^2_{1}}\Rb^{2{\bar \gamma}}
 \frac{4}{ ( 3 - 2 \,{\bar \gamma}) } \widetilde{G'}\Lb r'_\bot,\h
 Y - y_2 \Rb\,\widetilde{G'}\Lb r'_\bot,\h Y + y_2 \Rb\,\Bigg\} \,
\int d Q^2_T\, \, Q^4_T\,\nn\\
 &&\,N'^{\mbox{\tiny BK}}_i \Lb r_\bot,\h Y - y_1
 , Q_T\Rb\,N'^{\mbox{\tiny BK}}_j\Lb r_\bot,\h Y + y_1
 , Q_T\Rb\,N'^{\mbox{\tiny BK}}_i \Lb r'_\bot,\h Y -
 y_2 , Q_T\Rb\,N'^{\mbox{\tiny BK}}_j\Lb r'_\bot,\h
 Y + y_2 , Q_T\Rb\nn\eea
            
            and
    \bea \label{RCALD}
 &&D\Lb p_{1\bot}, p_{2\bot},Y, y_1,y_2\Rb\,\,=\,\,\\
 &&\sum^{i,j}_{i=1,j=1}\alpha^2_i \alpha^2_j \Bigg\{
 \int\frac{ d r^2_\bot }{r^4_\bot}J_0\Lb p_{1\bot}
 r_\bot\Rb\Lb\frac{ r^2_\bot}{r^2_{1}}\Rb^{2{\bar
 \gamma}}  \widetilde{G'}\Lb r_\bot,\h Y - y_1
 \Rb\,\widetilde{G'}\Lb r_\bot,\h Y + y_1 \Rb\Bigg\}\nn\\
 & \times&\,\, \Bigg\{ \int\frac{ d r'^2_\bot
 }{r'^4_\bot}J_0\Lb p_{1\bot} r'_\bot\Rb\Lb\frac{
 r'^2_\bot}{r^2_{1}}\Rb^{2{\bar \gamma}} \widetilde{G'}\Lb r'_\bot,\h
 Y - y_2 \Rb\,\widetilde{G'}\Lb r_\bot,\h Y + y_2 \Rb\,\Bigg\} \,\int
 d Q^2_T\, \, \,\nn\\
 &&\,N'^{\mbox{\tiny BK}}_i \Lb r_\bot,\h Y - y_1
 , Q_T\Rb\,N'^{\mbox{\tiny BK}}_j\Lb r_\bot,\h Y +
 y_1 , Q_T\Rb\,N'^{\mbox{\tiny BK}}_i \Lb r'_\bot,\h
 Y - y_2 , Q_T\Rb\,N'^{\mbox{\tiny BK}}_j\Lb r'_\bot,\h Y
 + y_2 , Q_T\Rb\nn\eea
            
 \section{Double inclusive cross section in the events with fixed multiplicity}
\begin{figure}[h]
\centerline{\epsfig{file=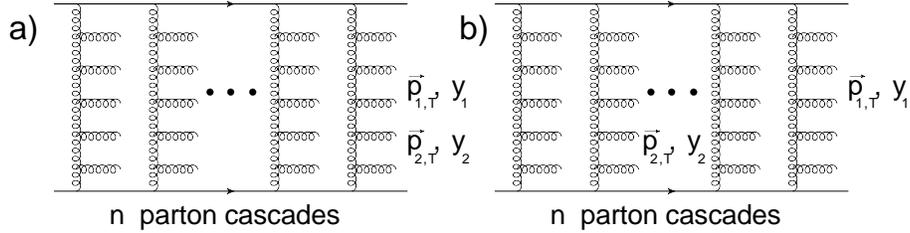,width=120mm}}
\caption{ The large multiplicity event in our approach: production from one parton  shower (\fig{n}-a) and from two parton showers ( \fig{n}-b). }
 \label{n} 
\end{figure}

 In this section we  calculate the angular dependance of
 the double inclusive cross section,  for the event with  large 
 multiplicity. In our approach, the large multiplicity event stems
 from the production of several parton showers, as it is shown in \fig{n}  
.
 Indeed, if  $N$ particles are produced in the collision,
 the $n$ parton showers contribute, where $n = N/\bar
 n$. $\bar n$ is the multiplicity in the single parton shower,
   which can 
be estimated as being equal to the average multiplicity in the single 
inclusive production.  Bearing this in mind, we  see that
 \bea\label{DI1}
 &&\frac{ d^2 \sigma}{d y_1 d y_2 d^2 p_{1T} d^2 p _{2T}}\,\,=\\
 &&\,\, n  \frac{ d^2 \sigma^{\mbox{\tiny one
 parton shower}}}{d y_1 d y_2 d^2 p_{1T} d^2 p _{2T}}  
 \,\,+\,\,n ( n - 1)\frac{ d^2 \sigma^{\mbox{\tiny two parton showers}}}{d
 y_1 d y_2 d^2 p_{1T} d^2 p _{2T}} \Big(
 1 \,+\,{\cal R}\Lb p_{1T},p_{2T}\Rb  \,\cos\Lb 2 \phi\Rb\Big)  \nn    
\eea
 
 The calculation of the first term in \eq{DI1} can be 
 simplified for $y_1 \approx y_2$ by using the following
 relation (see Ref.\cite{KLM})
 \beq \label{DI2}
  \frac{ d \sigma}{d y d^2 p_{T}} \,\,=\,\,\frac{8\,
 C_F}{\as \,(2 \,\pi)^4} \frac{1}{p^2_T} F^{\mbox{\tiny
 incl}}\Lb p_T\Rb;~~~\frac{ d^2 \sigma^{\mbox{\tiny one
 parton shower}}}{d y_1 d y_2 d^2 p_{1T} d^2 p _{2T}}|_{y_1 =
 y_2}\,\,=\,\,\Big( \frac{8\, C_F}{\as \,(2 \,\pi)^4} \Big)^2
 \,\frac{1}{p^2_{1T}\, p^2_{2T}}\, F^{\mbox{\tiny incl}}\Lb|\vec{p}_{1T}
 + \vec{p}_{2T}|\Rb  
  \eeq

The expression for the function $ F^{\mbox{\tiny incl}}$ we have found
 and it is equal to
\bea \label{DI3}
&& F^{\mbox{\tiny incl}}_{i,j}\Lb p_T,Y,y_1\Rb\,\,=\,\,\int\frac{
 d r^2_\bot }{r^4_\bot}J_0\Lb p_{1\bot} r_\bot\Rb\Lb\frac{
 r^2_\bot}{r^2_{1}}\Rb^{2{\bar \gamma}}\\
&&\times\,  \widetilde{G'}\Lb r_\bot,\h Y - y_1 \Rb\,\widetilde{G'}\Lb r_\bot,
\h Y + y_1 \Rb\,N'^{\mbox{\tiny BK}}_i \Lb r_\bot,\h Y - y_1 , Q_T=0\Rb\,N'^{
\mbox{\tiny BK}}_j\Lb r_\bot,\h Y + y_1 , Q_T=0\Rb\nn
\eea

Therefore, the first term in \eq{DI1}  reduces to the following
 expression
\beq \label{DI4}
n  \frac{ d^2 \sigma_{i,j}^{\mbox{\tiny one parton shower}}}{d
 y_1 d y_2 d^2 p_{1T} d^2 p _{2T}}\,\,=\,n\,\Big( \frac{8\,
 C_F}{\as \,(2 \,\pi)^4} \Big)^2 \,\frac{1}{p^2_{1T}\, p^2_{2T}}\,
 F_{i,j}^{\mbox{\tiny incl}}\Lb|\vec{p}_{1T} + \vec{p}_{2T}|\Rb  
\eeq
where $F^{\mbox{\tiny incl}}_{i,j}$ is given by \eq{DI3}. 

The second term in \eq{DI1} is almost equal to \eq{RCALD}
\bea \label{DI5}
&&n ( n - 1)\frac{ d^2 \sigma_{i,j}^{\mbox{\tiny two parton
 showers}}}{d y_1 d y_2 d^2 p_{1T} d^2 p _{2T}} \,\,=\,n\,\Lb n - 1\Rb 
  \Big(\frac{8\, C_F}{\as \,(2 \,\pi)^4} \Big)^2 \,\frac{1}{p^2_{1T}\,
 p^2_{2T}}\\\
&&\Bigg\{ \int\frac{ d r^2_\bot }{r^4_\bot}J_0\Lb p_{1\bot}
 r_\bot\Rb\Lb\frac{ r^2_\bot}{r^2_{1}}\Rb^{2{\bar \gamma}} 
 \widetilde{G'}\Lb r_\bot,\h Y - y_1 \Rb\,\widetilde{G'}\Lb
 r_\bot,\h Y + y_1 \Rb\Bigg\}\nn\\
 & \times&\,\, \Bigg\{ \int\frac{ d r'^2_\bot }{r'^4_\bot}J_0\Lb p_{1\bot}
 r'_\bot\Rb\Lb\frac{ r'^2_\bot}{r^2_{1}}\Rb^{2{\bar
 \gamma}} \widetilde{G'}\Lb r'_\bot,\h Y - y_2 \Rb\,\widetilde{G'}\Lb
 r_\bot,\h Y + y_2 \Rb\,\Bigg\} \,\int d Q^2_T\, \, \,\nn\\
 &&\,N'^{\mbox{\tiny BK}}_i \Lb r_\bot,\h Y - y_1 ,
 Q_T\Rb\,N'^{\mbox{\tiny BK}}_j\Lb r_\bot,\h Y + y_1 ,
 Q_T\Rb\,N'^{\mbox{\tiny BK}}_i \Lb r'_\bot,\h Y - y_2
 , Q_T\Rb\,N'^{\mbox{\tiny BK}}_j\Lb r'_\bot,\h Y - y_2
 , Q_T\Rb\nn\eea
 
 Using our calculation for inclusive production\cite{GLMINCL}, and for
 the rapidity correlation function we can re-write \eq{DI5} in the form
  \bea\label{DI1}
 &&\frac{1}{\sigma_{in}}\frac{ d^2 \sigma}{d y_1 d y_2 d^2 p_{1T}
 d^2 p _{2T}}\,\,=\,\,
 n \Big(\frac{8\, C_F}{\as \,(2 \,\pi)^4} \Big) \,\frac{1}{\sigma_{in}}
\frac{d \sigma}{d
 y_1}\frac{1}{p^2_{1T}\, p^2_{2
 T}}\, \frac{\sum^{i=2,j=2}_{i=1,j=1}\alpha^2_i\,
\alpha^2_j\,F_{i,j}^{\mbox{\tiny incl}}\Lb|\vec{p}_{1T} +
 \vec{p}_{2T}|\Rb  }{\sum^{i=2,j=2}_{i=1,j=1}\alpha^2_i\,
\alpha^2_j \int \frac{d^2 p_T}{p^2_T} F_{i,j}^{\mbox{\tiny
 incl}}\Lb p_T\Rb }\nn\\
 &&
   \,\,+\,\,\frac{n ( n - 1)}{2}\frac{1}{\sigma_{in}}\frac{d
 \sigma}{d y_1} \frac{1}{\sigma_{in}}\frac{d \sigma}{d y_2} 
\Big( 1 \,+\,R\Lb y_1,y_2\Rb\Big)\,\Big( 1\,+\,2 v_{22}\Lb p_{1T},p_{2T}\Rb
 \,\cos\Lb 2 \phi\Rb\Big)\,\nn\\
   &&\sum^{i=2,j=2}_{i=1,j=1}\alpha^2_i\,\alpha^2_j \,\, 
 \frac{d^2 \sigma_{i,j}^{\mbox{\tiny two parton showers}}}{d
 y_1 d y_2 d^2 p_{1T} d^2 p _{2T}}\Bigg{/}
   \sum^{i=2,j=2}_{i=1,j=1}\alpha^2_i\,\alpha^2_j \,\int d^2
 p_{1T}\, d^2 p_{2T}    \frac{d^2 \sigma_{i,j}^{\mbox{\tiny
 two parton showers}}}{d y_1 d y_2 d^2 p_{1T} d^2 p _{2T}}  
\eea 

 \section{Predictions and comparison with the experiment}
  \eq{DI1} has the form that has been used in the analysis of
 the experimental data\cite{ATLASACOR}: viz.
  \bea \label{PCE0}
&&\int d y_1 d y_2 \int^{p^{max}_{1T}}_{p^{min}_{1T}} d p_{1T}
 \int^{p^{max}_{2T}}_{p^{min}_{2T}}  d p_{2T} \frac{1}{\sigma_{in}}
\frac{ d^2 \sigma}{d y_1 d y_2 d p_{1T} d p _{2T} \,d \phi}  \equiv\,\\
&&\int d y_1 d y_2 \int^{p^{max}_{1T}}_{p^{min}_{1T}} d p_{1T}
 \int^{p^{max}_{2T}}_{p^{min}_{2T}}  d p_{2T} \frac{ d^2 N}{d y_1 d
 y_2 d p_{1T} d p _{2T} \,d \phi}  \,=\, Y^{\mbox{\tiny periph}}\Lb
 \phi\Rb\,\,+\,\,Y^{\mbox{\tiny ridge}}\Lb \phi\Rb\nn
\eea
where $\phi$ denotes the difference between the azimuthal angles
 $\phi = \phi_1 - \phi_2$.

$Y^{\mbox{\tiny periph}}\Lb \phi\Rb$ describes the production of two
 gluons from one parton shower (see \fig{n}-a )
 while   $Y^{\mbox{\tiny ridge}}\Lb \phi\Rb$ stands  for 
the
 second term, which is related to the emission of gluons from two 
different
 parton showers (see \fig{n}-b).
\eq{DI1} shows  several qualitative features which have been observed
 experimentally \cite{CMSACOR,ATLASACOR}:
 (1)     $Y^{\mbox{\tiny periph}}\Lb \phi\Rb$ is smaller  than
 $Y^{\mbox{\tiny ridge}}\Lb \phi\Rb$ and
 (2)   it decreases with  increasing  multiplicity
 of the event; (their ratio  is proportional to 1/n); and
 (3) $v_{22}$ does not depend on the multiplicity of the
 event and on the rapidity difference $Y_{12} = Y_1 - Y_2$.

From \eq{RCALN} and \eq{RCALD} we can see that
\bea \label{PCE1}
N \,&\propto&\int d^2 Q_T\,Q^4_T\, \Bigg( \int \frac{d ^2 r_\bot}{r^2_\bot}
 J_2\Lb p_{1T} r_{\bot}\Rb N_{i j}\Lb r_\bot, Q_T\Rb\Big)\Bigg( \int \frac{d
 ^2 r'_\bot}{r'^2_\bot} J_2\Lb p_{2T} r'_{\bot}\Rb N_{i j}\Lb r'_\bot,
 Q_T\Rb\Big);\nn\\
D &\propto& \int d^2 Q_T\,\, \Bigg( \int \frac{d ^2 r_\bot}{r^4_\bot}
 J_0\Lb p_{1T} r_{\bot}\Rb N_{i j}\Lb r_\bot, Q_T\Rb\Big)\Bigg( \int
 \frac{d ^2 r'_\bot}{r'^4_\bot} J_0\Lb p_{2T} r'_{\bot}\Rb N_{i j}\Lb
 r'_\bot, Q_T\Rb\Big);
\eea

\begin{figure}[h]
\centerline{\epsfig{file=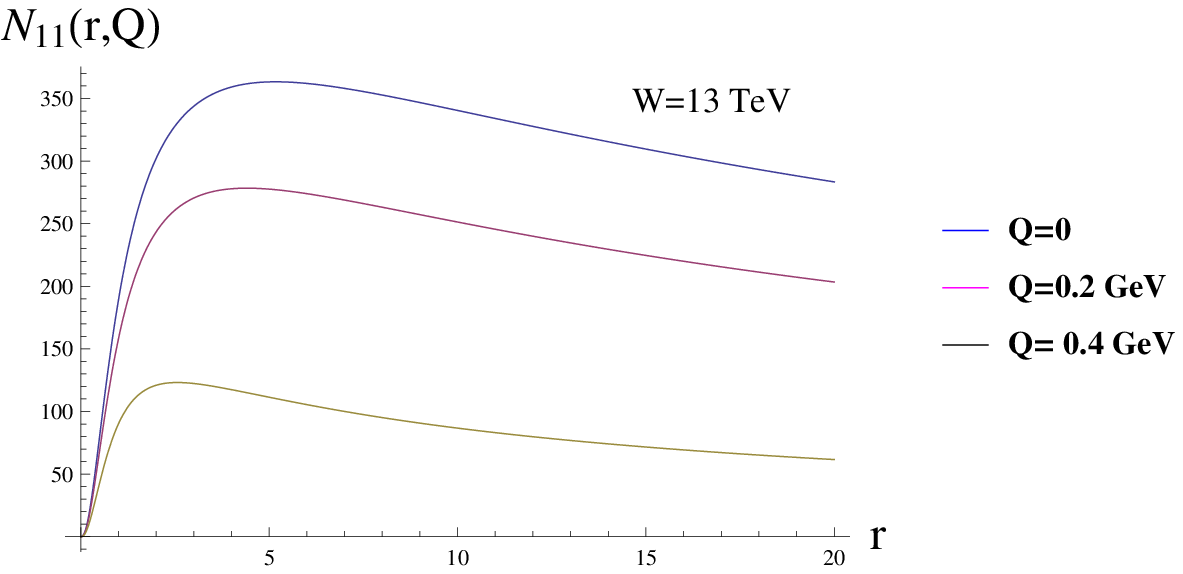,width=120mm}}
\caption{ $N_{11}$ of \eq{PCE1} versus $r_\bot$ at different values
 of $Q \equiv Q_T$ .}
 \label{n11} 
\end{figure}

As we have discussed, functions $N_{i, j}$ have maxima  at $r_\bot \approx
 1/Q_s$. In \fig{n11} we plot function $N_{11}$ at W = 13 TeV. One can see
 that $N_{11}$ has a maximum at $r_\bot \approx 3\div  4 \, GeV^{-1}$. 
Therefore,
  our estimates in section 4.2 are justified.

\begin{figure}[h]
\begin{center}
\begin{tabular}{c c}
\epsfig{file=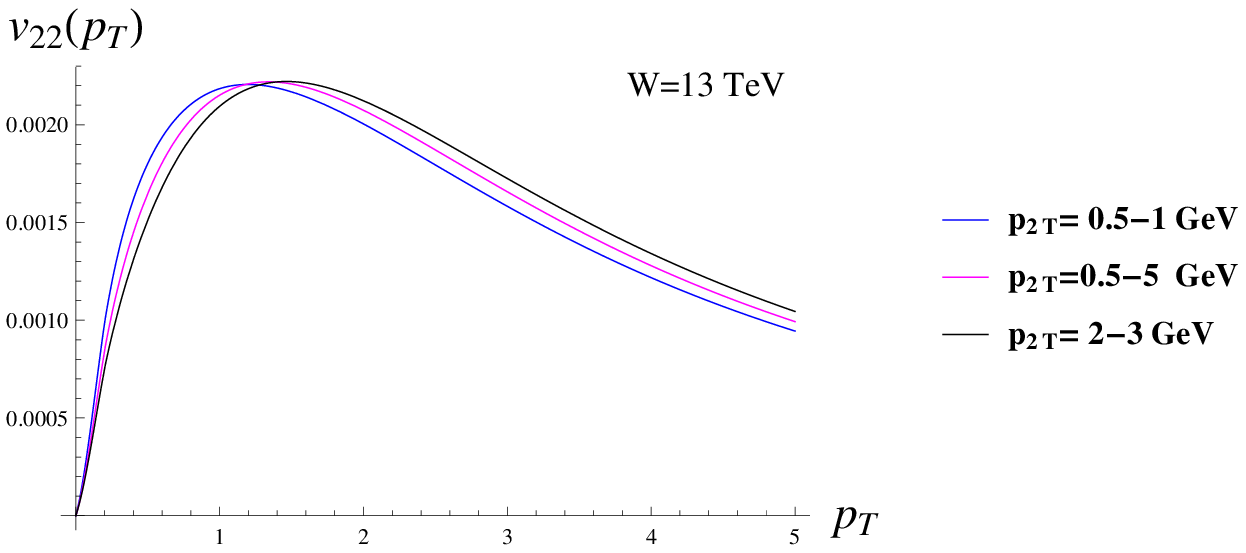,width=95mm}& \epsfig{file=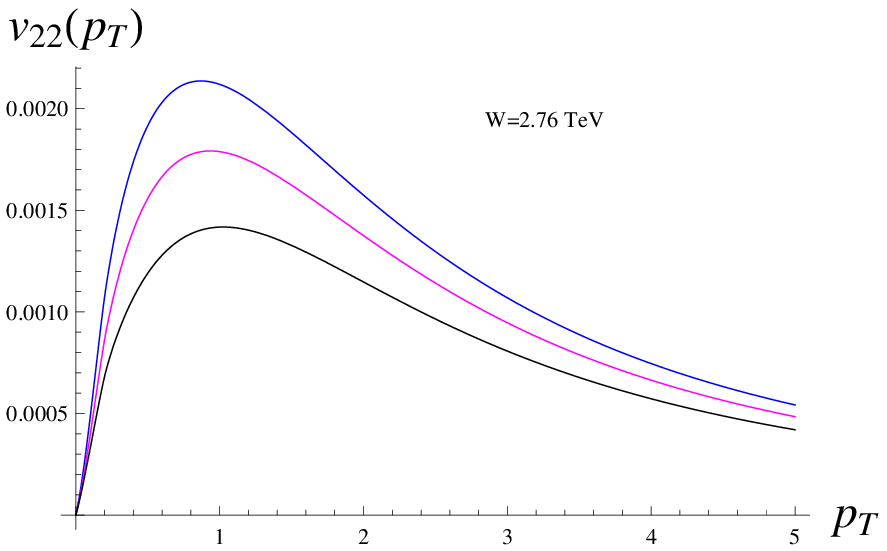,width=67mm}\\
\fig{v22lhc}-a & \fig{v22lhc}-b\\
\end{tabular}
\end{center}
\caption{ $\nu_{22}\, \equiv \,v_{22}$ versus $p_T$ at W = 13
 TeV ( \fig{v22lhc}-a)
 and at W = 2.76 TeV (\fig{v22lhc}-b).}
 \label{v22lhc} 
\end{figure}

The calculations of $v_{22} $ using 
\eq{RCAL} with the parameters of Table 1, are shown in \fig{v22lhc}.
 In \eq{PCE0} we integrate over $p_{2T}$ in the intervals shown in
 the figure:
 $ 0.5 \div 1 \, GeV$, $ 2 \div 3 \, GeV$ and $ 0.5 \div 5 \, GeV$.

\begin{figure}[h]
\begin{center}
\begin{tabular}{c c}
\epsfig{file=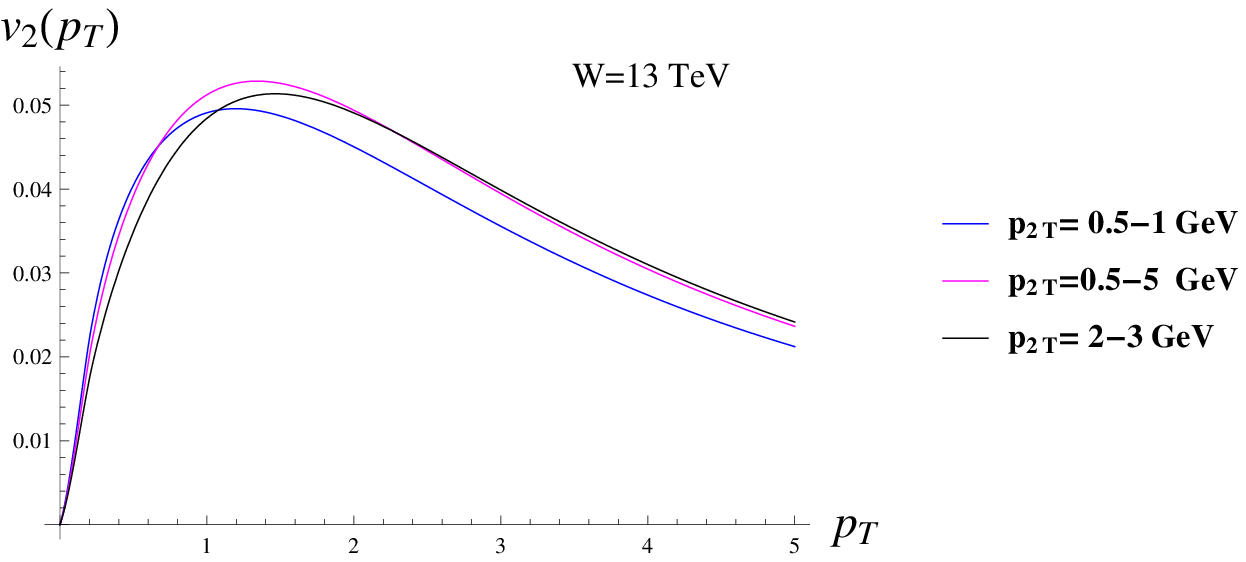,width=95mm}& \epsfig{file=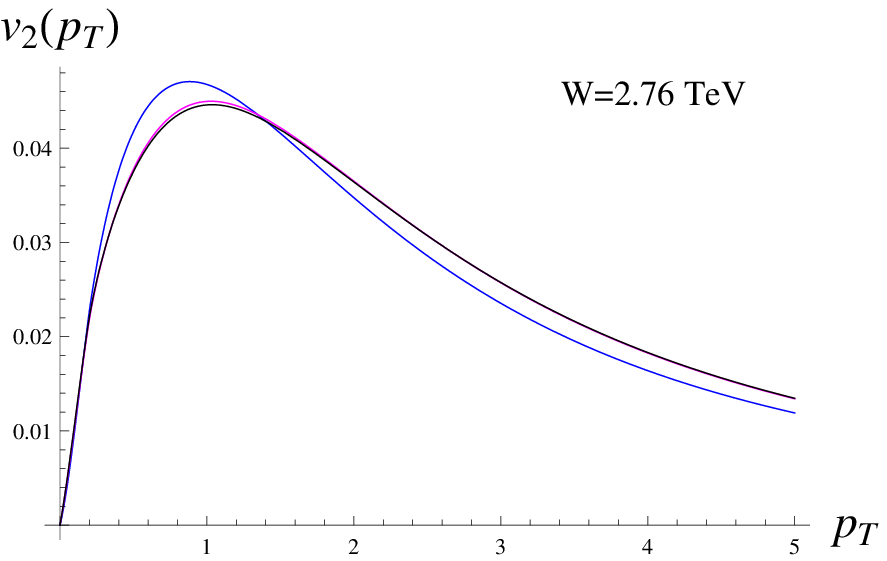,width=67mm}\\
\fig{v2lhc}-a & \fig{v2lhc}-b\\
\end{tabular}
\end{center}
\caption{ $\nu_2 \equiv v_{2}$ versus $p_T$ at W = 13 TeV (\fig{v2lhc}-a)
 and at W = 2.76 TeV (\fig{v2lhc} - b).}
 \label{v2lhc} 
\end{figure}

First conclusion from these figures is that $v_{22}$  
 does not  depend on energy, which is  in  agreement  with the 
experimental 
data of Ref.\cite{ATLASACOR}. 
. The values of $v_2(p_T) $  which is determined as 
\beq \label{PCE2}
v_2\Lb p_{1T} \Rb\,=\,\frac{v_{22}\Lb p_{1T},p_{2T}\Rb}{\sqrt{v_{22}\Lb
 p_{2T},p_{2T}\Rb }}
\eeq
are plotted in \fig{v2lhc}.

\begin{figure}[h]
\centerline{\epsfig{file=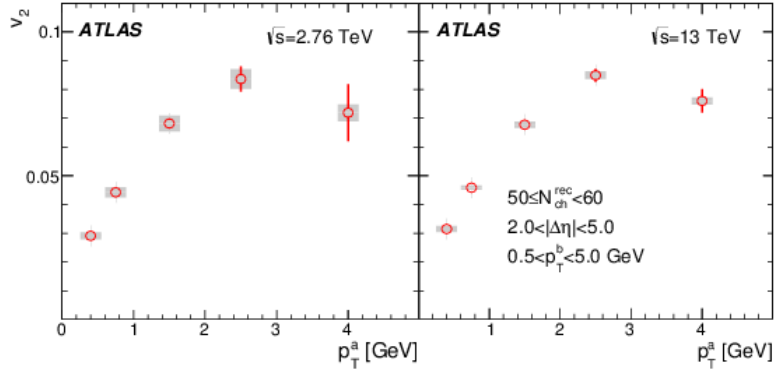,width=110mm}}
\caption{ $v_{2}$ versus $p_T$ at W = 2.76 TeV
 and at W = 13 TeV. The figure is taken from 
Ref.\cite{ATLASACOR}}
 \label{v2exlhc} 
\end{figure}

Comparing this figure with the experimental measurements of
 ATLAS \cite{ATLASACOR} we see:
(1) that the $p_T$ distribution
  for calculated $v_2$, is narrower than the experimental one; 
(2) it has a peak at $p_T =  0.5 \div 1.5 \, GeV$ while the experimental
  distribution has a maximum at $p_T =  2 \div \, 3 \, GeV$ and
 (3) the magnitude of  $ v_2$ is  about 0.6-0.7 of   the
 experimental one (see \fig{v2exlhc}).

\begin{figure}[h]
\begin{center}
\epsfig{file=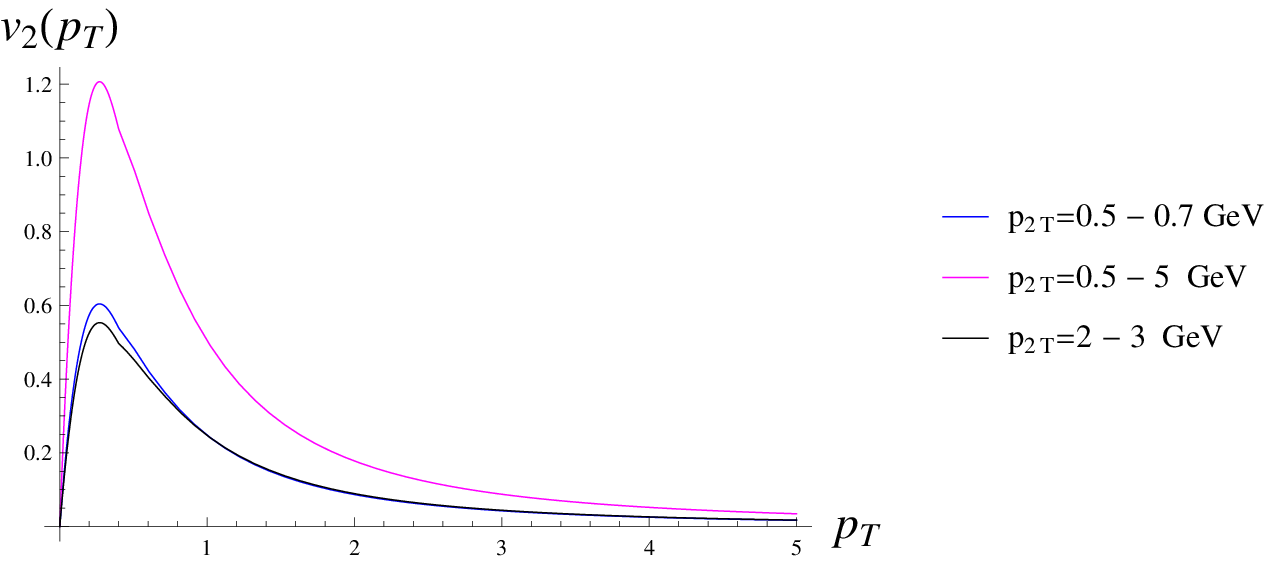,width=120mm}
\end{center}
\caption{ $v_{2}$ versus $p_T$  at W = 13 TeV with the
 choice $Q_0 = 0.2 m$ in \eq{NQ0}.}
 \label{v2nQs} 
\end{figure}

  In our calculation, we used \eq{QS0} for the
 saturation scale of \eq{QS}. This equation gives a natural
 generalization of our description of the diffractive processes
 using  only one  parameter, $m^2$. However, 
 we can introduce a different  scale:
\beq \label{NQ0}
Q^2_{0s}\Lb b, Y_0\Rb\,\, = \,\,Q^2_0 \Lb \frac{1}{m^2}
 S\Lb b, m\Rb\Rb^{\frac{1}{\bar \gamma}}
\eeq
 and view $Q_0$ as a new fitting parameter. 
In \fig{v2nQs} we plot the value of $v_2$ for $Q_0 = 0.2 m$.
  This choice does not change the character of
 $p_T$ dependence, but leads to a larger value than our previous 
estimates.

Therefore,  from comparison with the experiment data, we can conclude 
that
 the density variation mechanism in the framework of CGC/saturation
 approach, gives a substantial contribution which cannot be neglected.

   \section{Conclusion}
  In this paper we estimate the contribution of the density variation
 mechanism to the value of  $v_2$. This was done 
in a model based on
 CGC/saturation approach. It has been demonstrated in
 Refs.\cite{GLMNI,GLM2CH,GLMINCL,GLMCOR}, that the model is able to 
describe both
 the diffraction-type reactions ( the total and inelastic cross section,
 elastic cross section and its $t$ distribution, diffractive production)
 and the multi-particle production  
  processes, such as inclusive cross sections and rapidity correlations.
   We present here, the first attempt to describe  the azimuthal 
angular
 correlation, in the framework of a unique approach based on an 
effective
 theory for high energy QCD. 
  
Comparing with the experimental $v_2$ in proton-proton collision, we
 conclude  that the density variation mechanism in the framework of
 CGC/saturation approach, provides a  substantial contribution which
 cannot be neglected. Bearing in mind that  in CGC/saturation approach
 there are other two mechanisms  present: Bose enhancement in the 
wave
 function\cite{DDGJLR} and 
 local anisotropy\cite{KOLUCOR,KOLUREV}, we  believe that
  the azimuthal long range rapidity correlations in proton-proton
 collisions  stem from the CGC/saturation physics, and not from
  quark-gluon plasma production. It should be noted that none of
 the models based on the quark-gluon plasma production are able to
 describe diffractive physics. Hence, at present, the CGC/saturation 
approach,
  appears to be  the only effective theory that can 
provide such a description.
 
 We plan to include other CGC sources in the description of $v_2$, as
 well as compare in more detail, the cross section for double inclusive
 production
   of two gluon jets in proton-proton collisions.
   
  \section{Acknowledgements}
   We thank our colleagues at Tel Aviv university and UTFSM for
 encouraging discussions. Our special thanks go to   Gideon Bella and   
Carlos Cantreras for elucidating discussions on the
 subject of this paper. 
   This research was supported by the BSF grant   2012124 and  by  the
 Fondecyt (Chile) grant  1140842.


\begin{thebibliography}{99}
\bibitem{ALICE}
M.~G.~Poghosyan,
  J.\ Phys.\ G G {\bf 38}, 124044 (2011)
  [arXiv:1109.4510 [hep-ex]].
 ALICE ~Collaboration,
  {\it ``First proton--proton collisions at the LHC as observed with the ALICE
  detector: measurement of the charged particle pseudorapidity density at
  $\sqrt{s}$ = 900 GeV,''}
  arXiv:0911.5430 [hep-ex];\,\,\,
 A.~R.~Timmins [ALICE Collaboration],
  J.\ Phys.\ G {\bf 38} (2011) 124093;\,\,\,
 A.~R.~Timmins [ALICE Collaboration],
  arXiv:1106.6057 [nucl-ex];\,\,\,
  J.\ Phys.\ G {\bf 38} (2011) 124091
  [arXiv:1107.0285 [nucl-ex]].
\bibitem{ATLAS}
G.~Aad {\it et al.}  [ATLAS Collaboration],
  Nature Commun.\  {\bf 2},  463 (2011)
  [arXiv:1104.0326 [hep-ex]].
\bibitem{CMS}
CMS Physics Analysis Summary:
``Measurement of the inelastic pp cross section at $\sqrt{s}$ = 7  TeV with the CMS detector", 2011/08/27.

\bibitem{TOTEM}
 F.~Ferro [TOTEM Collaboration],
  AIP Conf.\ Proc.\  {\bf 1350}, 172 (2011) ;\,\,\,G.~Antchev {\it 
et al.}  [TOTEM Collaboration],
  Europhys.\ Lett.\  {\bf 96},  21002 (2011),
  {\bf 95},  41001 (2011)
  [arXiv:1110.1385 [hep-ex]];\,\,\,
G.~Antchev {\it et al.}  [TOTEM Collaboration],
  Phys.\ Rev.\ Lett.\  {\bf 111} (2013) 26,  262001
  [arXiv:1308.6722 [hep-ex]].
\bibitem{ALICEI}
ALICE Collaboration,
Eur.\ Phys.\ J.\  C {\bf 65} (2010) 111
[arXiv:0911.5430 [hep-ex]].
\bibitem{CMSI}
S.~Chatrchyan {\it et al.}  [CMS and TOTEM Collaborations],
  Eur.\ Phys.\ J.\ C {\bf 74} (2014) 10,  3053
  [arXiv:1405.0722 [hep-ex]];\,\,
   V.~Khachatryan {\it et al.}  [CMS Collaboration],
  JHEP {\bf 1002} (2010) 041
  [arXiv:1002.0621 [hep-ex]].
\bibitem{ATLASI}
ATLAS Collaboration,
arXiv:1003.3124 [hep-ex].
\bibitem{CMSMULT}
 V.~Khachatryan {\it et al.}  [CMS Collaboration],
  JHEP {\bf 1101} (2011) 079
  [arXiv:1011.5531 [hep-ex]].
\bibitem{ATLASCOR}
  G.~Aad {\it et al.} [ATLAS Collaboration],
  JHEP {\bf 1207} (2012) 019
  [arXiv:1203.3100 [hep-ex]].
\bibitem{KOLEB}
Yuri V Kovchegov and Eugene Levin, {\it `` Quantum Choromodynamics at High Energies"}, Cambridge Monographs on Particle Physics, Nuclear Physics and Cosmology, Cambridge University Press, 2012 .
\bibitem{GLMNI}
 E.~Gotsman, E.~Levin and U.~Maor,
  Eur.\ Phys.\ J.\ C {\bf 75} (2015) 1,  18
  [arXiv:1408.3811 [hep-ph]].
  \bibitem{GLM2CH}
 E.~Gotsman, E.~Levin and U.~Maor,
  Eur.\ Phys.\ J.\ C {\bf 75} (2015) 5,  179
  [arXiv:1502.05202 [hep-ph]].
\bibitem{GLMINCL}
 E.~Gotsman, E.~Levin and U.~Maor,
  Phys.\ Lett.\ B {\bf 746} (2015) 154
  [arXiv:1503.04294 [hep-ph]].
\bibitem{GLMCOR}
 E.~Gotsman, E.~Levin and U.~Maor,
 Eur.\ Phys.\ J.\ C {\bf 75} (2015) 11, 518
[arXiv:1508.04236 [hep-ph]].
\bibitem{GLMSP}
  E.~Gotsman, E.~Levin and U.~Maor,
  arXiv:1510.07249 [hep-ph].


\bibitem{GLR}
L. V. Gribov, E. M. Levin and M. G. Ryskin, 
Phys. Rep. {\bf 100} (1983) 1. 
\bibitem{MUQI}
A. H. Mueller and J. Qiu, 
Nucl. Phys. {\bf B268} (1986) 427.
\bibitem{MV}
L. McLerran and R. Venugopalan, 
Phys. Rev. {\bf D49} (1994) 2233, 3352; {\bf D50} (1994) 2225; 
{\bf D53} (1996) 458;\\ {\bf D59} (1999) 09400. 

\bibitem{B}
I.~Balitsky,
[arXiv:hep-ph/9509348];\,\,
{\it Phys.\ Rev.} {\bf D60}, 014020 (1999)
[arXiv:hep-ph/9812311].\,\,\,\,
\bibitem{MUCD}
A. H. Mueller,
Nucl. Phys. {\bf B415} (1994) 373; {\bf B437} (1995) 107.

\bibitem{K}
Y.~V.~Kovchegov,
{\it Phys.\ Rev.}  {\bf D60}, 034008  (1999),
[arXiv:hep-ph/9901281].
\bibitem{JIMWLK}
~J.~Jalilian-Marian, A.~Kovner, A.~Leonidov and H.~Weigert,
{\it  Phys.\ Rev.}\,  {\bf D59}, 014014 (1999),
[arXiv:hep-ph/9706377];\,\,  {\it Nucl.\ Phys.}\,{\bf B504}, 415
(1997),
[arXiv:hep-ph/9701284]; \,\,\,
J.~Jalilian-Marian, A.~Kovner and H.~Weigert,
  {\it Phys.\ Rev.}  {\bf D59}, 014015 (1999),
  [arXiv:hep-ph/9709432];\,\,\,
 A.~Kovner, J.~G.~Milhano and H.~Weigert,
 {\it  Phys.\ Rev.}  {\bf D62}, 114005 (2000),
  [arXiv:hep-ph/0004014]\,; \,\,\,
E.~Iancu, A.~Leonidov and L.~D.~McLerran,
{\it  Phys.\ Lett.}\,  {\bf B510}, 133 (2001);
[arXiv:hep-ph/0102009];\,\, {\it  Nucl.\ Phys.}\,  {\bf A692}, 583
(2001),
[arXiv:hep-ph/0011241];\,\,\,
E.~Ferreiro, E.~Iancu, A.~Leonidov and L.~McLerran,
 {\it  Nucl.\ Phys.}\  {\bf A703}, 489 (2002),
  [arXiv:hep-ph/0109115];\,\,\,
H.~Weigert,
{\it  Nucl.\ Phys.}  {\bf A703}, 823 (2002),
[arXiv:hep-ph/0004044].
%


\bibitem{BFKL}
 E. A. Kuraev, L. N. Lipatov, and F. S. Fadin, {\it  Sov. Phys.
JETP}
                {\bf 45}, 199 (1977); \,\,\,
Ya. Ya. Balitsky and L. N. Lipatov,
               {\it   Sov. J. Nucl. Phys.}\, {\bf 28}, 22 (1978).

\bibitem{LIREV} 
L. N. Lipatov,
Phys. Rep. {\bf 286} (1997) 131; Sov. Phys. JETP {\bf 63} (1986) 904 
and references therein. 




\bibitem{MUPA}
A. H. Mueller and B. Patel,
Nucl. Phys. {\bf B425} (1994) 471.

\bibitem{BART}
J. Bartels, M. Braun and G. P. Vacca,
Eur. Phys. J. {\bf C40} (2005) 419 [arXiv:hep-ph/0412218].
J. Bartels and C. Ewerz,
JHEP {\bf 9909} 026 (1999) [arXiv:hep-ph/9908454]. 
J. Bartels and M. Wusthoff,
Z. Phys. {\bf C6}, (1995) 157. 
J. Bartels,
Z. Phys. {\bf C60} (1993) 471.

 \bibitem{BRN}
M. A. Braun,
Phys. Lett. {\bf B632} (2006) 297 [arXiv:hep-ph/0512057]; 
Eur. Phys. J. {\bf C16} (2000) 337 [arXiv:hep-ph/0001268]; 
Phys. Lett. {\bf B483} (2000) 115 [arXiv:hep-ph/0003004]; 
Eur. Phys. J. {\bf C33} (2004) 113 [arXiv:hep-ph/0309293]; 
{\bf C6}, 321 (1999) [arXiv:hep-ph/9706373]. 
M. A. Braun and G. P. Vacca,
Eur. Phys. J. {\bf C6} (1999) 147 [arXiv:hep-ph/9711486].

\bibitem{KOLE}
Y.~V.~Kovchegov and E.~Levin,
  Nucl.\ Phys.\ B {\bf 577} (2000) 221
  [hep-ph/9911523].
 

\bibitem{LELU}
E. Levin and M. Lublinsky,
  Nucl.\ Phys.\  A {\bf 763} (2005) 172
  [arXiv:hep-ph/0501173];\,\,
  Phys.\ Lett.\  B {\bf 607} (2005) 131
  [arXiv:hep-ph/0411121];\,\, 
  Nucl.\ Phys.\  A {\bf 730} (2004) 191
  [arXiv:hep-ph/0308279].
\bibitem{LMP}
E. Levin, J. Miller and A. Prygarin,
  Nucl.\ Phys.\  {\bf A806 } (2008)  245,
  [arXiv:0706.2944 [hep-ph]].
\bibitem{AKLL}
  T.~Altinoluk, C.~Contreras, A.~Kovner, E.~Levin, M.~Lublinsky and A.~Shulkim,
  Int.\ J.\ Mod.\ Phys.\ Conf.\ Ser.\  {\bf 25} (2014) 1460025;\,\,\,
   T.~Altinoluk, N.~Armesto, A.~Kovner, E.~Levin and M.~Lublinsky,
  JHEP {\bf 1408} (2014) 007;\,\,\,
 T.~Altinoluk, A.~Kovner, E.~Levin and M.~Lublinsky,
  JHEP {\bf 1404} (2014) 075
  [arXiv:1401.7431 [hep-ph]].;\,\,\, 
  T.~Altinoluk, C.~Contreras, A.~Kovner, E.~Levin, M.~Lublinsky and A.~Shulkin,
  JHEP {\bf 1309} (2013) 115.
\bibitem{LEPP}
E.~Levin,
  JHEP {\bf 1311} (2013) 039
  [arXiv:1308.5052 [hep-ph]].
\bibitem{KLN}
D.~Kharzeev, E.~Levin and M.~Nardi,
  Nucl.\ Phys.\  A {\bf 730} (2004) 448
  [Erratum-ibid.\  A {\bf 743} (2004) 329]
  [arXiv:hep-ph/0212316];\,\
  Phys.\ Rev.\  C {\bf 71} (2005) 054903
  [arXiv:hep-ph/0111315];\,\, D.~Kharzeev and E.~Levin,
  Phys.\ Lett.\  B {\bf 523} (2001) 79
  [arXiv:nucl-th/0108006];\,\,\, D.~Kharzeev and M.~Nardi,
  Phys.\ Lett.\  B {\bf 507} (2001) 121
  [arXiv:nucl-th/0012025].



\bibitem{KLNLHC}
  D.~Kharzeev, E.~Levin and M.~Nardi,
  Nucl.\ Phys.\  A {\bf 747} (2005) 609
  [arXiv:hep-ph/0408050].

\bibitem{LERE}
 E.~Levin and A.~H.~Rezaeian,
  AIP Conf.\ Proc.\  {\bf 1350} (2011) 243
  [arXiv:1011.3591 [hep-ph]];\,\,
  Phys.\ Rev.\ D {\bf 82} (2010) 014022
  [arXiv:1005.0631 [hep-ph]]; \,\, 
[arXiv:1102.2385 [hep-ph]];\,\,
{\it Phys. Rev} {\bf D82} (2010) 054003.
[arXiv:1007.2430 [hep-ph]].

\bibitem{LERECOR}
E.~Levin and A.~H.~Rezaeian,
  Phys.\ Rev.\ D {\bf 84}, 034031 (2011)
  [arXiv:1105.3275 [hep-ph]].



\bibitem{MCLPR}
L. McLerran, M. Praszalowicz,
  Nucl.\ Phys.\ A {\bf 916} (2013) 210
  [arXiv:1306.2350 [hep-ph]];\,\,
{\it Acta Phys. Polon.} {\bf B42 } (2011) 99 
[arXiv:1011.3403 [hep-ph]];\,\,
{\it Acta Phys. Polon.} {\bf B41} (2010) 1917.
[arXiv:1006.4293 [hep-ph]].
\bibitem{PRA}
M. Praszalowicz,
  Phys.\ Lett.\ B {\bf 727} (2013) 461
  [arXiv:1308.5911 [hep-ph]];\,\,
  Acta Phys.\ Polon.\ Supp.\  {\bf 6} (2013) 3,  809
  [arXiv:1304.1867 [hep-ph]];\,\,
  Phys.\ Lett.\ B {\bf 704} (2011) 566
  [arXiv:1101.6012 [hep-ph]]l\,\,
  Phys.\ Rev.\ Lett.\  {\bf 106} (2011) 142002
  [arXiv:1101.0585 [hep-ph]].







\bibitem{COL}
P.D.B. Collins, {\it "An introduction to Regge theory and high energy physics"}, 
Cambridge University Press 1977.

\bibitem{MUDI}
A. H. Mueller,
{\it Phys. Rev.} {\bf D2} (1970) 2963.
\bibitem{CMSPP}
  V.~Khachatryan {\it et al.}  [CMS Collaboration],
  JHEP {\bf 1009} (2010) 091
  [arXiv:1009.4122 [hep-ex]].

\bibitem{STARAA}
  J.~Adams {\it et al.}  [STAR Collaboration],
  Phys.\ Rev.\ Lett.\  {\bf 95} (2005) 152301
  [nucl-ex/0501016].
\bibitem{PHOBOSAA}
  B.~Alver {\it et al.}  [PHOBOS Collaboration],
  Phys.\ Rev.\ Lett.\  {\bf 104} (2010) 062301
  [arXiv:0903.2811 [nucl-ex]].
\bibitem{STARAA1}
  H.~Agakishiev {\it et al.}  [STAR Collaboration],
  arXiv:1010.0690 [nucl-ex].
\bibitem{CMSPA}
 S.~Chatrchyan {\it et al.}  [CMS Collaboration],
  Phys.\ Lett.\ B {\bf 718} (2013) 795
  [arXiv:1210.5482 [nucl-ex]].
\bibitem{CMSAA}
  S.~Chatrchyan {\it et al.}  [CMS Collaboration],
  Eur.\ Phys.\ J.\ C {\bf 72} (2012) 2012
  [arXiv:1201.3158 [nucl-ex]].

\bibitem{CAUSALITY}
A.~Dumitru, F.~Gelis, L.~McLerran and R.~Venugopalan,
  Nucl.\ Phys.\ A {\bf 810} (2008) 91
  [arXiv:0804.3858 [hep-ph]].
\bibitem{FINSTATE}
E.~V.~Shuryak,
  Phys.\ Rev.\ C {\bf 76} (2007) 047901
  [arXiv:0706.3531 [nucl-th]];\,\,\,S.~A.~Voloshin,
  Phys.\ Lett.\ B {\bf 632} (2006) 490
  [nucl-th/0312065];\,\,\,S.~Gavin, L.~McLerran and G.~Moschelli,
  Phys.\ Rev.\ C {\bf 79} (2009) 051902
  [arXiv:0806.4718 [nucl-th]].
\bibitem{KOVT}
A. Kovner, {\it ``How to build a ridge in  pA
collisions"}, talk at Low x WS, May 30 - June 4, 2013,Rehovot-Eilat, Israel.~~$
http://www.weizmann.ac.il/MaKaC/getFile.py/access?contribId=60\&sessionId=25\&resId=0\&materialId=slides\&confId=16$

\bibitem{DDGJLR}
  A.~Dumitru, K.~Dusling, F.~Gelis, J.~Jalilian-Marian, T.~Lappi and R.~Venugopalan,
  Phys.\ Lett.\ B {\bf 697} (2011) 21
  [arXiv:1009.5295 [hep-ph]].


\bibitem{KOLUCOR}
A.~Kovner and M.~Lublinsky,
  Phys.\ Rev.\ D {\bf 83} (2011) 034017
  [arXiv:1012.3398 [hep-ph]].

\bibitem{KOLUREV}
A.~Kovner and M.~Lublinsky,
  Int.\ J.\ Mod.\ Phys.\ E {\bf 22} (2013) 1330001
  [arXiv:1211.1928 [hep-ph]].

\bibitem{REV1}
 W.~Li,
  Mod.\ Phys.\ Lett.\ A {\bf 27} (2012) 1230018
  [arXiv:1206.0148 [nucl-ex]].
\bibitem{REV2}
A.~Kovner,
  Acta Phys.\ Polon.\ B {\bf 42} (2011) 2717.
\bibitem{REV3}
C.~J.~Horowitz,
  Int.\ J.\ Mod.\ Phys.\ E {\bf 20} (2011) 1
  [arXiv:1106.1661 [astro-ph.SR]].
\bibitem{REV4}
 T.~Lappi,
  Int.\ J.\ Mod.\ Phys.\ E {\bf 20} (2011) 1
  [arXiv:1003.1852 [hep-ph]].
\bibitem{REV5}
E.~Iancu,
  arXiv:1205.0579 [hep-ph].
\bibitem{REV6}
R.~Venugopalan,
  PoS QNP {\bf 2012} (2012) 019
  [arXiv:1208.5731 [hep-ph]].
  
  \bibitem{MPSI}
A. H. Mueller and B. Patel,
Nucl. Phys. {\bf B425} (1994) 471. 
A. H. Mueller and G. P. Salam,
Nucl. Phys. {\bf B475}, (1996) 293. [arXiv:hep-ph/9605302]. 
G. P. Salam,
Nucl. Phys. {\bf B461} (1996) 512; 
E. Iancu and A. H. Mueller,
Nucl. Phys. {\bf A730} (2004) 460 [arXiv:hep-ph/0308315];            
494 [arXiv:hep-ph/0309276].




\bibitem{MUTR}
A.~H.~Mueller and D.~N.~Triantafyllopoulos,
{\it Nucl.\ Phys.} \, {\bf B640} (2002) 331
[arXiv:hep-ph/0205167];
\bibitem{GW}
M. L. Good and W. D. Walker, Phys. Rev. 120 (1960) 1857.


\bibitem{MUPE}
S.~Munier and R.~B.~Peschanski,
  Phys.\ Rev.\  D {\bf 70} (2004) 077503
  [arXiv:hep-ph/0401215];\,\,
Phys.\ Rev.\  D {\bf 69} (2004) 034008
  [arXiv:hep-ph/0310357];\,\,
  Phys.\ Rev.\ Lett.\  {\bf 91} (2003) 232001
  [arXiv:hep-ph/0309177].
\bibitem{NAPE}
~H.~Navelet ~ and ~R.~B.~ Peschanski,  Nucl. Phys. {\bf B 507}, 35 (1997) [hep-ph/9703238];\,
  Phys.\ Rev.\ Lett.\  {\bf 82} (1999) 1370
  [hep-ph/9809474];\,\,Nucl.\ Phys.\ B {\bf 634} (2002) 291
  [hep-ph/0201285].
\bibitem{RY}
I. Gradstein and I. Ryzhik, {\it  Table of Integrals, Series, and Products},
Fifth Edition, Academic Press, London, 1994.

\bibitem{IIMU}
E.~Iancu, K.~Itakura and S.~Munier,
  Phys.\ Lett.\ B {\bf 590}, 199 (2004)
  [hep-ph/0310338].

\bibitem{KTF}
S.~Catani, M.~Ciafaloni and F.~Hautmann,
  Nucl.\ Phys.\ B {\bf 366} (1991) 135;
 Nucl. Phys. Proc. Suppl. 29A (1992) 182;\,\,\,J.~C.~Collins and R.~K.~Ellis,
  Nucl.\ Phys.\ B {\bf 360} (1991) 3;\,\,\,
E.~M.~Levin, M.~G.~Ryskin, Y.~.M.~Shabelski and A.~G.~Shuvaev,
  Sov.\ J.\ Nucl.\ Phys.\  {\bf 53} (1991) 657
   [Yad.\ Fiz.\  {\bf 53} (1991) 1059].

  \bibitem{KTINC}
   Y.~V.~Kovchegov and K.~Tuchin,
  Phys.\ Rev.\   {\bf D65} (2002) 074026
  [arXiv:hep-ph/0111362].



\bibitem{DPI}
G.~Aad {\it et al.}  [ATLAS Collaboration],
  New J.\ Phys.\  {\bf 15} (2013) 033038
  [arXiv:1301.6872 [hep-ex]],\,\,\,
 V.~M.~Abazov {\it et al.}  [D0 Collaboration],
  Phys.\ Rev.\ D {\bf 81} (2010) 052012
  [arXiv:0912.5104 [hep-ex]],\,\,\,
 F.~Abe {\it et al.}  [CDF Collaboration],
  Phys.\ Rev.\ D {\bf 47} (1993) 4857,\,\,\,
  Phys.\ Rev.\ D {\bf 56} (1997) 3811,\,\,
J.~Alitti {\it et al.}  [UA2 Collaboration],
  Phys.\ Lett.\ B {\bf 268} (1991) 145,\,\,\,
T.~Akesson {\it et al.}  [Axial Field Spectrometer Collaboration],
  Z.\ Phys.\ C {\bf 34}, 163 (1987).
\bibitem{KNO}
Z. Koba, H. B. Nielsen, and P. Olesen, 
Nucl. Phys. B40 (1972) 317.
\bibitem{V2PP}
 E.~Avsar, C.~Flensburg, Y.~Hatta, J.~Y.~Ollitrault and T.~Ueda,
  Phys.\ Lett.\ B {\bf 702}, 394 (2011)
  doi:10.1016/j.physletb.2011.07.031
  [arXiv:1009.5643 [hep-ph]].
\bibitem{CMSACOR}
  V.~Khachatryan {\it et al.} [CMS Collaboration],
  JHEP {\bf 1009} (2010) 091
  doi:10.1007/JHEP09(2010)091
  [arXiv:1009.4122 [hep-ex]].

\bibitem{ATLASACOR}
 G.~Aad {\it et al.} [ATLAS Collaboration],
  {\it ``Observation of long-range elliptic anisotropies in $\sqrt{s}=$13 and 2.76 TeV $pp$ collisions with the ATLAS detector,''}
  arXiv:1509.04776 [hep-ex].

\bibitem{KLM}
D.~Kharzeev, E.~Levin and L.~McLerran,
  Nucl.\ Phys.\ A {\bf 748} (2005) 627,
  [hep-ph/0403271].

%



  \end{thebibliography}
\end{document}